\documentclass[a4paper,onecolumn,10pt, accepted=2022-07-10]{quantumarticle}
\pdfoutput=1

\usepackage[utf8]{inputenc}
\usepackage[english]{babel}
\usepackage[T1]{fontenc}
\usepackage{amsmath,amssymb,amsthm,mathrsfs}
\usepackage{hyperref}
\usepackage[numbers,sort&compress]{natbib}

\usepackage{bbm}
\usepackage{xcolor}
\usepackage{multirow}
\usepackage{graphicx}%
\usepackage{soul,color}

\usepackage{tikz}
 
\usepackage{pgfplots}
\usepackage{pifont}

\newcommand{\bra}[1]{\langle #1|}
\newcommand{\ket}[1]{| #1 \rangle}

\newcommand{\cprnd}{$\mathsf{cprnd}$}
\newcommand{\lrs}{$\mathsf{lrs}$ }
\newcommand{\vinci}{$\mathsf{vinci}$ }

\renewcommand{\S}{\mathcal{S}}
\newcommand{\T}{\mathcal{T}}
\newcommand{\NS}{\mathcal{NS}}
\newcommand{\M}{\mathcal{M}}
\newcommand{\Mp}{\mathcal{M}^\text{\tiny P}}
\newcommand{\Qc}{\Q_{[\cdot,\cdot]}}
\newcommand{\Qt}{\Q_{**}}

\newcommand{\sP}{\mathscr{P}}
\newcommand{\Q}{\mathcal{Q}}
\newcommand{\AQ}{\widetilde{\Q}}

\newcommand{\dd}{{\rm d}}
\newcommand{\dSe}{\dd s_\text{\tiny E}}
\newcommand{\dSf}{\dd s_\text{\tiny F}}
\newcommand{\dVe}{\dd V_\text{\tiny E}}
\newcommand{\dVf}{\dd V_\text{\tiny F}}
\newcommand{\RVe}[1]{\text{RV}(#1)}
\newcommand{\tp}{^\text{\tiny T}}

\newcommand{\ns}{n_s}
\newcommand{\no}{n_o}

\newcommand{\Nt}{N_\text{\scriptsize tot}}
\newcommand{\tr}{\ensuremath{\operatorname{tr}}}

\newcommand{\Id}{\mathbb{I}_d}
\newcommand{\vecP}{\vec{P}}

\DeclareMathOperator*{\argmin}{\text{\small argmin}}

\newcommand{\De}{\mathfrak{D}_\text{\tiny E}}

\newcommand{\Dnpal}{D_{\Q_k}}
\newcommand{\Dml}{D_{\AQ_\ell}}
\newcommand{\Dmesl}{D_{\Mp_h}}

\renewcommand{\L}{\mathcal{L}}
\renewcommand{\P}{\mathcal{P}}
\newcommand{\unit}{{100.00$^{**}$\%} }
\usepackage[capitalize]{cleveref}

\newtheorem{Observation}{Observation}

\begin{document}
\title{Naturally restricted subsets of nonsignaling correlations: typicality and convergence}

\author{Pei-Sheng Lin}
\affiliation{Department of Physics and Center for Quantum Frontiers of Research \& Technology (QFort), National Cheng Kung University, Tainan 701, Taiwan}
\author{Tam\'as V\'ertesi}
\affiliation{MTA Atomki Lend\"ulet Quantum Correlations Research Group, Institute for Nuclear Research, P.O. Box 51, H-4001 Debrecen, Hungary}
\author{Yeong-Cherng Liang}
\email{ycliang@mail.ncku.edu.tw}
\affiliation{Department of Physics and Center for Quantum Frontiers of Research \& Technology (QFort), National Cheng Kung University, Tainan 701, Taiwan}
\affiliation{Physics Division, National Center for Theoretical Sciences, Taipei 10617, Taiwan}

\begin{abstract}
It is well-known that in a Bell experiment, the observed correlation between measurement outcomes---as predicted by quantum theory---can be stronger than that allowed by local causality, yet not fully constrained by the principle of relativistic causality. In practice, the characterization of the set $\Q$ of quantum correlations is carried out, often, through a converging hierarchy of {\em outer} approximations.  On the other hand, some subsets of $\Q$  arising from additional constraints [e.g., originating from quantum states having positive-partial-transposition (PPT) or being finite-dimensional maximally entangled (MES)] turn out to be also amenable to similar numerical characterizations. How,  then, at a {\em quantitative} level, are all these naturally restricted subsets of nonsignaling correlations different? Here, we consider several bipartite Bell scenarios and numerically estimate their volume relative to that of the set of nonsignaling correlations.  Within the number of cases investigated, we have observed that (1) for a given number of inputs $\ns$ (outputs $\no$), the relative volume of both the Bell-local set and the quantum set increases (decreases) rapidly with increasing $\no$ ($\ns$) (2)  although the so-called macroscopically local set $\Q_1$ may approximate $\Q$ well in the two-input scenarios, it can be a very poor approximation of the quantum set when $\ns>\no$  (3) the almost-quantum set $\AQ_1$ is an exceptionally-good approximation to the quantum set  (4) the difference between $\Q$ and the set of correlations originating from MES is most significant when $\no=2$, whereas (5) the difference between the Bell-local set and the PPT set generally becomes more significant with increasing $\no$. This last comparison, in particular, allows us to identify Bell scenarios where there is little hope of realizing the Bell violation by PPT states and those that deserve further exploration.
\end{abstract}
\maketitle

\section{Introduction}
\label{sec:Introduction}

The fact that Bell inequalities~\cite{Bell64}---constraints derived from the assumption of Bell-locality~\cite{Brunner_RevModPhys_2014}---can be violated by quantum theory indicates that the set of quantum correlations  $\Q$ is intrinsically different from the set of correlations $\L$ allowed by a locally-causal theory~\cite{Bell04}. However, it is also known---from the pioneering work of Popescu and Rohrlich~\cite{Popescu_PRBox_1994}---that quantum theory is {\em not} the most Bell-nonlocal (hereafter abbreviated as nonlocal) among all physical theories that respect the principle of relativistic causality. In  bipartite Bell scenarios, this principle gives rise to the so-called nonsignaling (NS) conditions~\cite{Barrett_05}, and hence a superset of $\Q$ known as the  nonsignaling polytope $\NS$.
 
Since then, a lot of effort (see, e.g.,~\cite{Brassard_NCCT,Linden_LNC,Pawlowski_IC,Navascues_MacroscopicLocal,Fritz_LO,Navascues_AlmostQuantum,Gonda2018almostquantum}) has been devoted to understand if {\em additional} physical, or information-theoretic principles can be supplemented to recover from $\NS$ the set of quantum correlations. In fact, the so-called {\em almost-quantum} set of correlations~\cite{Navascues_AlmostQuantum}---known to be a strict superset to $\Q$---apparently satisfies all information-theoretic principles proposed to date. Nonetheless, the extent to which this set, hereafter denoted by $\AQ_1$, differs from $\Q$ itself is not well understood. Indeed, \cite{Vallins2017} seems to be the only work to date reporting a systematic investigation of the difference in Bell values achievable by these sets. 

So far, the only known means that we have in characterizing $\Q$ is via a hierarchy of outer approximations, such as that proposed by Navascu\'es, Pironio, and Ac\'in (NPA)~\cite{NPA}. The NPA hierarchy is known to converge~\cite{NPA2008} (see also~\cite{Doherty08})---in the asymptotic limit---to the set of correlations $\Qc$ achievable assuming quantum theory and with the measurements between spatially separated parties modeled by commuting operators, rather than tensor products. Note that the two different formulations of spatially separated measurements generally lead to different~\cite{SLOFSTRA:2019aa,SLOFSTRA:2020,Coladangelo:2020wn} sets of correlations, i.e., $\Q\neq \Qc$, thereby manifesting the complication of the geometry of these sets~\cite{Goh2018}. Moreover, very little is known~\cite{Harrow2019} regarding the rate of convergence of this hierarchy towards $\Qc$.

On the other hand, a few other {\em subsets} of $\Q$ are naturally also of interest. For example, a somewhat different formulation~\cite{Moroder13} of the NPA hierarchy has made it possible to characterize---also via a converging hierarchy of outer approximations---the set of correlations $\P$ arising from quantum states having positive partial transposition~\cite{Peres96} (PPT). The interest in this stems from a conjecture of Peres~\cite{Peres:1999aa}---disproved in~\cite{Vertesi_PPTState}---concerning the impossibility of bound entangled~\cite{Horodecki1998} states violating Bell inequalities. Known counterexamples to Peres' conjecture are, however, too fragile to be demonstrated in any experiment, thus making it desirable to understand how $\P$, being a {\em restricted} subset of $\Q$, differs from the set of Bell-local correlations $\L$.

Besides, the fact that certain nonlocal features only seem to exist for partially entangled states is also intriguing. One of the first hints along this line is the Hardy paradox~\cite{Hardy1993}. Later, the existence of such correlations was explicitly shown, independently, in~\cite{Liang2011,Vidick2011,Junge:2011aa} (see also~\cite{Christensen15}), under the name of {\em more nonlocality with less entanglement}. Interestingly, the set of correlations $\M$ arising from finite-dimensional maximally entangled states, or more precisely its convex hull---as with $\Q$ and $\P$---can also be characterized~\cite{Lang_MES} via a hierarchy of outer approximations, each corresponds to the feasible set of a semidefinite program~\cite{BoydBook}. To achieve a better understanding of the precise relationship between entanglement and nonlocality, any quantitative estimate of the difference between $\M$ and $\Q$ is surely welcome.

Apart from fundamental interests, nonlocal correlations also play an indispensable role in the context of device-independent quantum information~\cite{Scarani_DIQI_12,Brunner_RevModPhys_2014}. For instance, from the observation of a Bell violation itself, one can certify the generation of unpredictable random bits~\cite{Colbeck2006,Pironio10}, guarantee the sharing of truly unconditional secured keys~\cite{Acin07}, certify various desired features of the underlying systems (see, e.g.,~\cite{Bancal11,Moroder13,Liang:PRL:2015,SLChen16,Arnon-Friedman:2019aa,Chen_18}), measurements (see, e.g.,~\cite{Rabelo2011,SLChen16,Bancal:PRL:2018,Chen:PRR:2021}) or even other more general types of operations~\cite{Sekatski2018,Wagner2020}. In its strongest form, one could  achieve so-called self-testing~\cite{Mayers04}, where the underlying system and measurements are identified uniquely, modulo unimportant local degrees of freedom. For a comprehensive review on this last topic, see~\cite{Supic19}.

To this end, it is worth noting that the quantitative differences between the ``size" of the Bell-local set $\L$, the quantum set $\Q$, and the nonsignaling set $\NS$ has been investigated in~\cite{Cabello_05,Elie_12,Duarte_18}. Specifically, in the simplest Bell scenario involving two parties, each performing two binary-outcome measurements, the volume of $\L$ and that of $\Q$, relative to $\NS$, in the {\em subspace} of ``full" correlation functions~\cite{Werner:PRA:2001} was first determined in~\cite{Cabello_05}. Then, for the same Bell scenario, the analysis has been generalized~\cite{Elie_12} to include also the subspace spanned by marginal correlations. Beyond this, numerical estimation of the relative volume of $\L$ to $\NS$ was carried out in~\cite{Duarte_18} for a few Bell scenarios with either {\em only} two measurement settings or outcomes; some analytic results were also presented therein when restricted to the subspace of full correlations.

Here, we generalize the analysis of~\cite{Cabello_05,Elie_12,Duarte_18} in two directions.  Firstly, we consider a more extensive list of Bell scenarios, including a few with multiple measurement settings and outcomes, which allows us to make observations that were not possible in prior works. Secondly, we consider not only $\L$, $\Q$ (approximated by relevant outer approximations), and $\NS$, but also the set of correlations $\P$ achievable by PPT quantum states, the (convex hull of the) set of correlations achievable by locally measuring finite-dimensional maximally entangled states with projective measurements $\Mp$, the set $\Q_1$ associated with the principle of {\em macroscopic locality}~\cite{Navascues_MacroscopicLocal},  the almost-quantum set $\AQ_1$, and more generally the first few levels of the outer approximations of $\Q$ given by the NPA hierarchy~\cite{NPA} as well as the hierarchy of Moroder {\em et al.}~\cite{Moroder13}. This last consideration, in particular, allows us to learn the fraction of points in $\NS$ that can be further excluded from $\Q$ as we consider increasingly tighter outer approximations to $\Q$. Clearly, such information is highly relevant for device-independent analyses that rely on the above hierarchies to outer-approximate $\Q$.

The rest of this paper is structured as follows. In ~\cref{Sec:Preliminaries}, we introduce the notations used throughout the paper and the different Bell scenarios considered. Then, we present in~\cref{Sec:RV-Euclidean} our main results on the relative volume for the different sets of correlations mentioned above. In ~\cref{sec:Discussion}, we give further discussions  and comment on some possible directions for future research. 

\section{Preliminaries}
\label{Sec:Preliminaries}

\subsection{Notations and Naturally Restricted Subsets of $\NS$}

Consider a bipartite Bell experiment where each spatially separated party has a choice over $\ns$ measurement settings where each measurement results in $\no$ possible outcomes. We shall denote such a Bell scenario by ($\ns,\no$). Correlation between the observed outcomes for given measurement settings of the two parties (conventionally called Alice and Bob) may be described by $\vecP = \{P(a,b|x,y)\}_{a,b,x,y}$ where we label the measurement settings and outcomes for Alice (Bob), respectively, as $x$ ($y$) and $a$ ($b$).  Throughout, we consider only Bell scenarios where $x,y,a$, and $b$ take a finite number of values.

Our starting point is the set of nonsignaling correlations~\cite{Popescu_PRBox_1994,Barrett_05}, $\NS$, which are all those $\vecP$ that satisfy the so-called nonsignaling conditions~\cite{Barrett_05}:
\begin{equation}
\label{eq:NS}
\begin{split}
	\sum_{a} P(a,b|x,y) \overset{\NS}{=} \sum_{a} P(a,b|x',y)\;\; \forall \;b,y,x,x',\\
	\sum_{b} P(a,b|x,y) \overset{\NS}{=} \sum_{b} P(a,b|x,y')\;\; \forall \;a,x,y,y'.
\end{split}
\end{equation}
Physically, these conditions were initially~\cite{Popescu_PRBox_1994} proposed to exclude the possibility of communication by making different local choices of $x,y$. Since $\NS$ is the intersection of a finite number of hyperplanes defined by~\cref{eq:NS} and the direct sum of $\ns^2$ probability simplices, it is a convex polytope.

An important subset of $\NS$ is the Bell-local~\cite{Brunner_RevModPhys_2014} set $\L$. For any  $\vecP\in\L$, it can be shown~\cite{Fine.PRL.1982,Pitowsky:Book} (see also Ref.~\cite{Wiseman:2014vo}) that there exists normalized weights $q(\lambda) \ge\; 0$ for all $\lambda$ and  $\sum_\lambda q(\lambda) = 1$ such that 
\begin{equation}
\begin{split}
  \label{eq:LHV}
  &P(a,b|x,y)  \overset{\L}{=} \sum_\lambda q(\lambda)\delta_{a,f_A(x,\lambda)}\delta_{b,f_B(y,\lambda)},\\
\end{split}
\end{equation}
for some choice of local response functions $f_A(x,\lambda)$ and $f_B(y,\lambda)$. 
Clearly, $\L$ is convex. Since there are only finite possibilities of $x,y,a$, and $b$, the set $\L$ forms~\cite{Pitowsky:Book} a convex polytope, with its extreme points correspond to local deterministic strategies given by the Kronecker deltas: $\delta_{a,f_A(x,\lambda)}$ and $\delta_{b,f_B(y,\lambda)}$.

Suppose now that the two parties share a  quantum state $\rho$, then quantum theory dictates that the observed correlation follows Born's rule:
\begin{equation}\label{eq:Q}
	P(a,b|x,y) \overset{\Q}{=} {\rm tr}\left[\rho\, M^{(A)}_{a|x}\otimes M^{(B)}_{b|y}\right], 
\end{equation}
where $\{M^{(A)}_{a|x}\}_{a}$ $\left(\{M^{(B)}_{b|y}\}_{b}\right)$ are positive-operator-valued measures (POVMs) associated with Alice's (Bob's) $x$-th ($y$-th) measurement. We denote the set of such correlations by $\Q$.  Importantly, in the definition of $\Q$, there is no constraint imposed on the (local) Hilbert space dimension. This, in turn, guarantees the convexity~\cite{Pitowsky:Book} of $\Q$ and the sufficiency of projective measurements (via Naimark's extension~\cite{Peres:1990aa}) on pure state $\rho=\rho^2$ in the membership test of $\Q$. Even then, the characterization of $\Q$ is  by no means  computationally easy.

To this end, NPA~\cite{NPA} first pointed out that $\Q$, or more precisely, $\Qc$ can be characterized~\cite{NPA2008} {\em asymptotically} (see also~\cite{Doherty08}) by solving a hierarchy of semidefinite programs (SDPs). The feasible sets corresponding to these SDPs then define a series of outer approximations $\Q_1\supseteq\Q_2\supseteq\cdots\supseteq\Q_\infty=\Qc\supset\Q$, where $\Q_k$ is referred conventionally as NPA level $k$ with $k$ signifying the highest degree of the operator used in defining some matrix of moments (see~\cite{NPA,NPA2008} for details). $\Q_1$, incidentally, is exactly the set of correlations that respects the principle of macroscopic locality~\cite{Navascues_MacroscopicLocal}. Intermediate levels can also be considered and a prominent example is the so-called NPA level $1+AB$, which happens to be the lowest level of another converging hierarchy of outer approximations $\AQ_1\supseteq\AQ_2\supseteq\cdots\supseteq\AQ_\infty=\Qc\supset\Q$ due to Moroder {\em et al.}~\cite{Moroder13}. The specific outer approximation $\Q_{1+AB}=\AQ_1$ is known in the literature as the {\em almost quantum}~\cite{Navascues_AlmostQuantum} set of correlations, as it seems to satisfy all principles that have been proposed to date to distinguish $\Q $ from $\NS$. In the bipartite scenario, the sets associated with the two hierarchies are known~\cite{Moroder13} to satisfy the inclusion relations $\Q_1 \supset \AQ_1$ and $\AQ_k \supset  \Q_{2k}$ for all integers $k\ge1$.

As Bell showed in his seminal work~\cite{Bell64}, there exist quantum correlations arising from entangled quantum states that do not admit a convex decomposition in the form of~\cref{eq:LHV}. At the same time, by considering trivial POVMs consisting only of the identity operator and the null operator, it is straightforward to see that all $\vecP\in\L$ can always be cast in the form of \cref{eq:Q}. Similarly, all $\vecP\in\Q$ are easily seen to satisfy ~\cref{eq:NS}, while Popescu and Rohrlich~\cite{Popescu_PRBox_1994} showed that there exists $\vecP\in\NS\setminus\Q$. Together, one arrives at the strict inclusion relations $\L \subsetneq \Q \subsetneq \AQ_{1}\subsetneq \Q_1\subsetneq \NS$. 

Even if we restrict our attention to $\Q$ itself, its relationship with the set of quantum states has not been fully understood. For example, although entanglement is necessary~\cite{Werner:PRA:1989} for Bell-nonlocality, some entangled states (see, e.g., ~\cite{Werner:PRA:1989, Barrett:PRA:2002}) are known to produce {\em only} $\vecP\in\L$. In fact, even if Alice and Bob are allowed to share an arbitrary finite-dimensional maximally entangled state $\ket{\Psi_d} = \frac{1}{\sqrt{d}}\sum_{i=1}^{d}\ket{i}\ket{i}$, it is {\em impossible} for them to reproduce all $\vecP\in\Q$~\cite{Liang2011,Vidick2011,Junge:2011aa,Christensen15}. To facilitate subsequent discussions, we denote by $\Mp$  the convex hull of the set of correlations attainable by performing projective measurements on $\ket{\Psi_d}$, with $d$ finite. On the contrary, even the weakest form of entanglement given by PPT entangled states may generate nonlocal correlations~\cite{Vertesi:PRL:2012,Vertesi_PPTState}. Hereafter, we denote by $\P$ the subset of $\Q$ which arises from $\rho$ being a PPT state, i.e., $\rho^{\text{\tiny T}_A} \succeq 0$ where $^{\text{\tiny T}_A}$ stands for the partial transposition operation on Alice's Hilbert space. 

\begin{figure}
  \centering
  \includegraphics[scale=0.15]{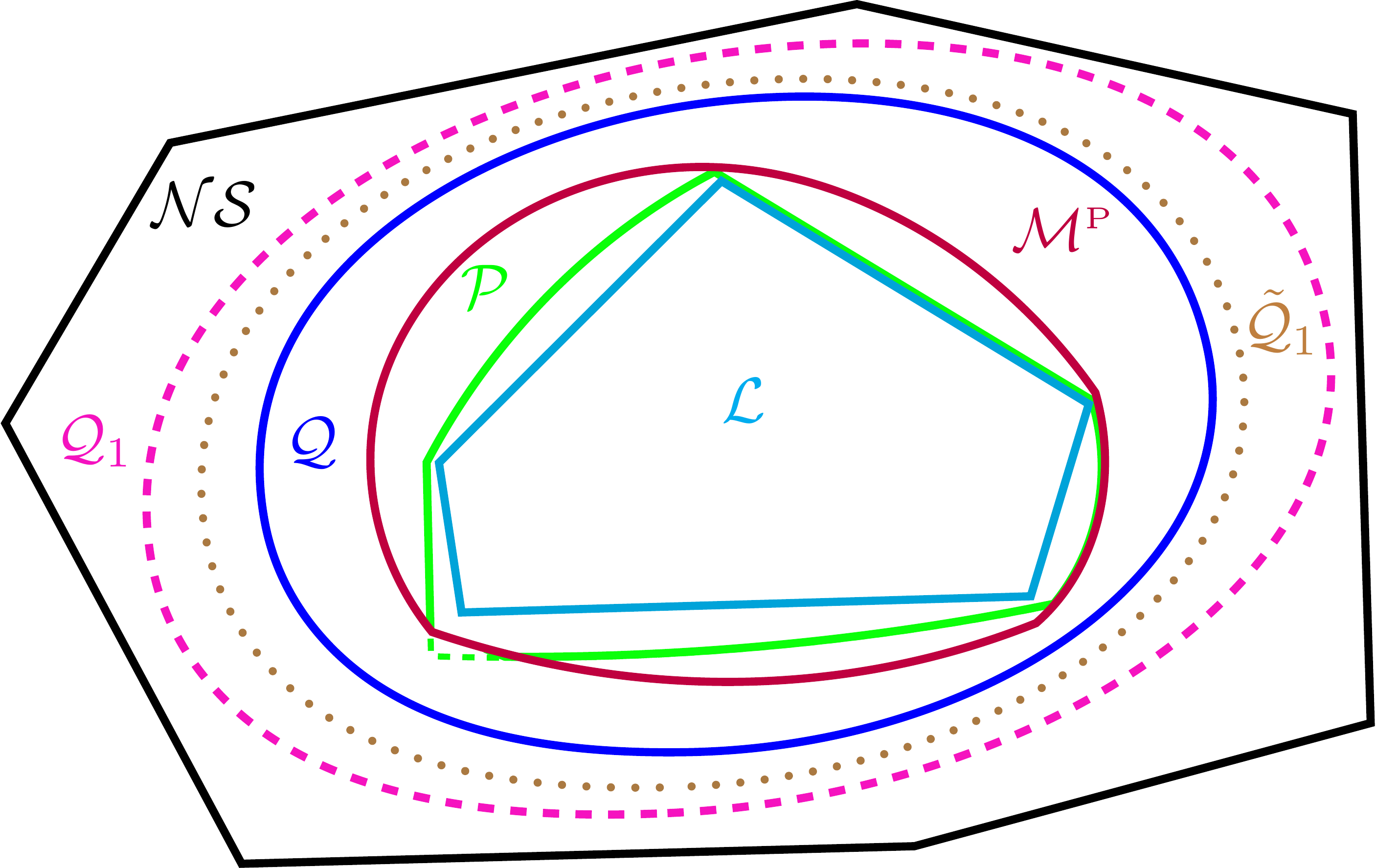}
  \caption{\label{fig:RelationOfSets} Schematic illustration of the relationships among the various subsets $\T\subset \NS$ considered here. 
Starting from $\NS$ and moving inwards, we have in solid lines, respectively, the boundary of the quantum set of correlations $\Q$ [\cref{eq:Q}, blue], the convex hull $\Mp$ (red) of the set of correlations attainable by finite-dimensional maximally entangled states in conjunction with projective measurements, the set of correlations attainable by PPT entangled state $\P$ (green), and the Bell-local polytope $\L$ (skyblue). Dashed (pink) and dotted (brown) lines lying between the boundary of $\Q$ and that of $\NS$ mark the boundary of the lowest-level outer approximation of $\Q$, respectively, due to NPA~\cite{NPA} ($\Q_1$) and Moroder {\em et al.}~\cite{Moroder13} ($\AQ_1$). 
}
\end{figure}

A schematic diagram explaining the relationships among the various naturally restricted subsets $\T$ of $\NS$ considered in this work is provided in~\cref{fig:RelationOfSets}. 

\subsection{Membership Tests}
\label{sec:membership}

Our goal is to estimate the relative volume of various subsets $\T\in\{\L, \P, \Mp, \Q, \AQ_1, \Q_1\} $ of $\NS$. To this end, we perform the membership test $\vecP\overset{?}{\in}\T$ 
for each sampled $\vecP\in\NS$ by solving the following optimization problem:
\begin{subequations}
  \label{eq:Visibility}
  \begin{align}
    \sup\;\; &v \label{Eq:obj} \\
    {\rm s.t.}\;\; v\vecP +(1-&v)\vecP_w \in \T, \label{Eq:membership} 
  \end{align}
\end{subequations}
where $\vec{P}_w$ is the uniform probability distribution, i.e., $P_w(a,b|x,y) = 1/\no^2\; \forall\; x,y$. As $\vecP_w$ lies strictly in $\L$, it must also lie in {\em all} sets $\T$  that are of our interest. Hence, the above optimization problem, which we solve using the optimization software MOSEK implemented in MATLAB, is always feasible by setting $v=0$. Also, if $\vecP \in \T$ then  all mixtures with $v \in [0,1]$ are inside $\T$, i.e., the optimum $v$ (denoted by $v^*$) would be greater than or equal to 1. Hence, $v^* < 1$ indicates that $\vec{P} \not\in \T$. Notice that $v^*$, often called the white-noise visibility, can be understood as the ``maximal" weight that can be assigned to $\vecP$ when it is admixed with white noise while ensuring that the mixture lies within $\T$. A smaller value of $v^*$, which corresponds to a larger value of $1-v^*$, then indicates that the correlation is more robust (in terms of preserving its nonlocal nature) against the mixing with $\vecP_w$. 

Among the different sets of interest, $\L$ is a convex polytope, and thus its membership test, cf.~\cref{eq:Visibility} is an instance of a linear program~\cite{BoydBook}. For relatively simple Bell scenarios, this optimization problem can be efficiently solved on a computer. In contrast, for the other sets of interest, including $\T\in\{\P, \Mp, \Q\}$, we  rely on a hierarchy of outer approximations, each of which is amenable to semidefinite programming characterizations. In the case of $\Q$, we use both the NPA hierarchy~\cite{NPA,NPA2008} and its variant due to Moroder {\em et al.}~\cite{Moroder13} for membership tests. For definiteness, we denote by $\Q_k$ and $\AQ_{\ell}$, respectively, the level $k$ and the level $\ell$ outer approximation of $\Q$ based on the NPA hierarchy and the hierarchy of Moroder {\em et al.} (a summary of both hierarchies can be found in Table V, Appendix B of Ref.~\cite{SLChen18}). By further requiring the moment matrix  $\AQ_{\ell}$ to be PPT, one immediately obtains a characterization of $\P_\ell$, i.e., the level $\ell$ outer approximation of $\P$. Notice that all these SDPs can be implemented using the $\mathsf{Ncpol2sdpa}$  toolbox developed by Wittek~\cite{Wittek}. For the set of correlations $\Mp$ associated with local projective measurements on finite-dimensional maximally entangled states, we make use of a hierarchy adapted from that presented in Ref.~\cite{Lang_MES}, the details of which are given in~\cref{app:RelaxationSDP}. We refer to the level $h$ outer approximation of $\Mp$ obtained thereof as $\Mp_h$.

\subsection{Metrics and Relative Volume}
\label{sec:RV}

The notion of volume for any given region in a space $\sP$ is metric-dependent. In our case, $\sP$ is the set of conditional probability distributions $\vecP = \{P(a,b|x,y)\}_{a,b,x,y}$ where $a,b\in\{1,\ldots, \no\}$ and $x,y\in\{1,2,\ldots, \ns\}$. The normalization requirement $\sum_{a,b} P(a,b|x,y)=1$ for all $x$ and $y$ implies that $\mathscr{P}$ is $(\ns^2)(\no^2-1)$-dimensional. Moreover, we are only interested in $\vecP$ that satisfy the nonsignaling constraints of \cref{eq:NS}. The nonsignaling polytope $\NS$ and hence the various subsets of interest all lie in a $d$-dimensional subspace $\sP_{\NS}$ of $\sP$ where~\cite{Collins04} $d=(\no-1)^2\ns^2+2\ns(\no-1)$.

A convenient, {\em minimal} parametrization of any $\vecP\in\NS$ is given by~\cite{Collins04}: 
\begin{equation}\label{Eq:CG}
	\vecP = \{P(a|x), P(b|y),P(a,b|x,y)\}_{a,b,x,y}
\end{equation}
where $P(a|x) = \sum_b P(a,b|x,y)$ and $P(b|y) = \sum_a P(a,b|x,y)$ are, respectively, the marginal conditional probability distributions of Alice and Bob. Note that in this parameterization, the labels $a,b$ in \cref{Eq:CG} take only values from $\{1,\dots,\no-1\}$. Indeed, the conditional probability distributions for the omitted outcome corresponding to $a$ and/or $b=\no$ can be determined easily from the components $P_i$ of $\vecP$ in \cref{Eq:CG} via the normalization of probabilities and the nonsignaling conditions of ~\cref{eq:NS}. 

Like in previous works~\cite{Cabello_05,Elie_12,Duarte_18}, we adopt the Euclidean metric in our computation of the relative volumes (though other options may also be considered, see our remark in~\cref{sec:Discussion}). In this metric $\dSe^2$, all components of $\vecP$ in \cref{Eq:CG} are treated on equal footing. Explicitly, $\dSe^2$ and the corresponding volume element $\dVe$ are given, respectively, by:
\begin{equation}\label{eq:EuclideanMetric}
    \dSe^2 = \sum_{i} {\dd} P_i^2\quad \text{ and }\quad
    \dVe = \prod_{i}{\dd} P_i. 
\end{equation}
When there is no risk of confusion, the subscript $_\text{E}$ is omitted to simplify the presentation.

We then define the relative volume (RV) for each set $\T\subset\NS$ as:
\begin{equation}
	\RVe{\T}\equiv \frac{V(\T)}{V(\NS)}
\end{equation}
where $V(\S)$ is the volume of a set $\S$ in accordance to the (Euclidean) metric. To numerically estimate these RVs, it suffices to sample points $\vecP\in\NS$ \emph{uniformly} according to the metric, and determine the fraction of such points that lie in $\T$ via the method explained in \cref{sec:membership}. For this purpose, we make use of the MATLAB function \cprnd~ developed by Benham~\cite{cprnd}, and in particular its \emph{Gibbs sampler} algorithm to perform uniform sampling of $\vecP$ in $\NS$. In~\cref{app:SamplingMethod}, we give further details on how we generate uniform samples in $\NS$ using the \cprnd~function. Note that each membership test with respect to $\T$ is exactly a Bernoulli trial with a success probability given by $\RVe{\T}$.

For each Bell scenario considered, we estimate the RV of $\L$, $\Q_l$, $\AQ_k$, $\P_k$ and $\Mp_h$. In~\cref{Tbl:Results_Detail_Euclidean}, we list all the bipartite Bell scenarios considered in this work, the number of samples used in estimating the RV of each target set,  the highest level of each type of hierarchies considered, as well as the corresponding RVs. In ~\cref{app:Scenarios-Complexity}, we further provide the relevant parameters characterizing the size of the optimization problem. In the next section, we present our main results across 23 Bell scenarios.

For all but two of these Bell scenarios, we use $\Nt=10^6$ sampled correlations  in our estimation of the relative volumes. The only exceptions are the (5,4) and the (6,3) Bell scenario where we employ, instead, $3.05\times10^5$ and $7.85\times10^5$ samples, respectively. Even in these latter cases, the estimates have converged well with the employed samples (see~\cref{fig.Conv.ns.sample} in~\cref{app.Convergence} for details). Importantly, as we can see in these plots, the number of samples required to reach a good precision does {\em not} seem to depend on the complexity of the Bell scenario, but rather more on the success probability itself.

\section{Numerical Estimates of Relative Volumes}
\label{Sec:RV-Euclidean}  

\begin{table*}[t!]
    \includegraphics[width=1\linewidth]{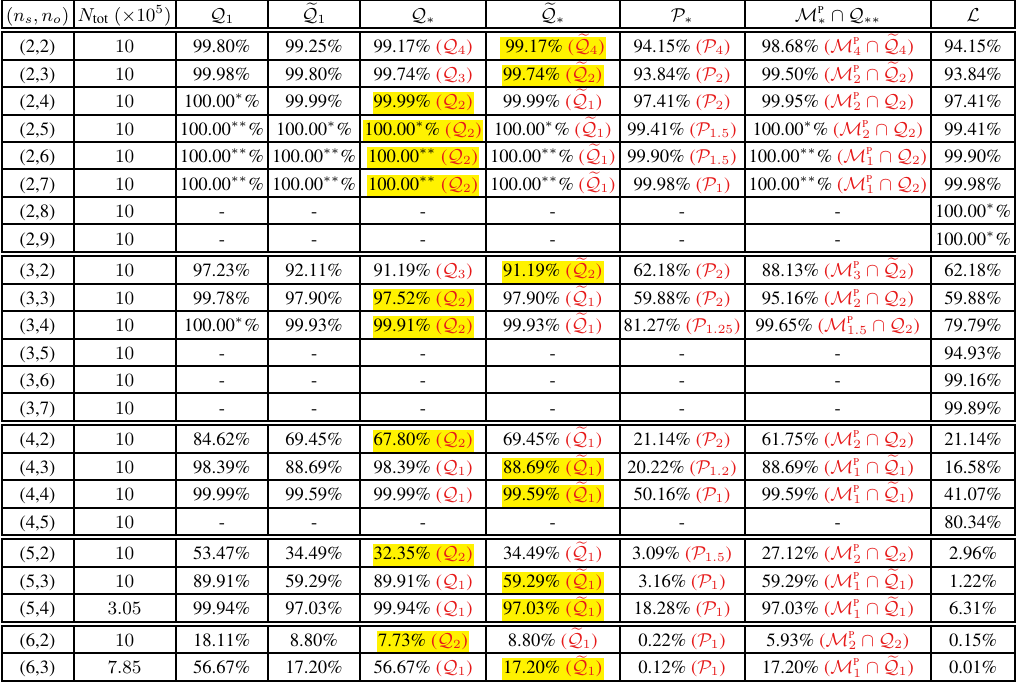}
  \caption{\label{Tbl:Results_Detail_Euclidean} Summary of the numerically estimated relative volume $\RVe{\T}$ for various naturally restricted subsets $\T$ of the set of nonsignaling correlations $\NS$. The second column gives the number of $\vecP$ uniformly sampled from $\NS$ using \cprnd. From the third column to the rightmost column, we have the estimated relative volume (RV) for, respectively,  the macroscopically local set $\Q_1$, the almost quantum set $\AQ_1$,  our tightest approximation to $\Q$ based on outer approximations of NPA~\cite{NPA} (denoted by $\Q_*$),  our tightest approximation to $\Q$ based on outer approximations of Moroder~{\em et al.}~\cite{Moroder13} (denoted by$\AQ_*$), our tightest approximation to $\P$ (denoted by $\P_*$), our tightest approximation to $\Mp$ intersecting with $\Qt$ (denoted by $\Mp_*\cap \Qt$), and the Bell-local set $\L$. In the fifth to the eighth column, we include also  in bracket the highest level of the SDP hierarchy used in the computation (for an explanation of the various levels and the complexity involved in the computation, see~\cref{app:Scenarios-Complexity}). In particular, our best approximation to $\RVe{\Q}$, given either in the fifth or the sixth column, is highlighted in yellow. For example, in our characterization of $\Q$ in the $(2,3)$ Bell scenario, we are not able to go beyond $\Q_3$ nor $\AQ_2$, neither of which is, {\em a priori}, a subset of the other. However, since $\AQ_2$ gives a smaller RV (see~\cref{tbl:ConvergenceResults_Euclidean}), we use it as our $\Qt$ in this Bell scenario.  We use 100.00$^*$\%  and \unit to denote entries where the {\em estimated}  $\RVe{\T}$ satisfies, respectively, $\RVe{\T}> 99.995\%$ and $\RVe{\T}> 100\% \left(1- \frac{1}{\Nt}\right)$. Similarly, all other estimates reported here have a fundamental imprecision of $\frac{100}{\Nt}\%$.  Here and below, entries marked with ``-" means the corresponding computation has been left out.}
\end{table*}

In~\cref{Tbl:Results_Detail_Euclidean}, we provide a summary of $\RVe{\T}$ for $\T\in\{\Q_1,\AQ_1,\L\}$ and certain approximations to $\Q$, $\P$, and $\Mp$ for the various Bell scenarios considered. Note that for 6 of the 23 Bell scenarios, we compute only $\RVe{\L}$ as it becomes too time consuming to compute the other $\RVe{\T}$ with a statistically significant number of trials. Throughout, we use $\Qt$, $\Mp_*$, and $\P_*$, respectively, to denote the tightest approximation that we are able to compute for $\Q$, $\Mp$, and $\P$. In the following subsections, we describe in details how the RV of these sets changes in different Bell scenarios. 
To best illustrate these trends, we make use of line plots showing how each of these relative volumes varies with respect to the relevant parameters.

\subsection{$\L$ vs $\NS$}
\label{Sec:LvsNS}

\begin{figure}[h!]
  \centering
  \includegraphics[scale=1]{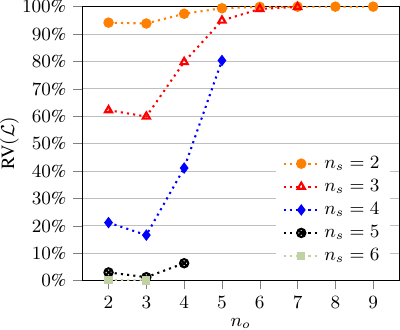}
\caption{\label{fig:Euclidean:LvsNS} Plots of estimated $\RVe{\L}$ vs $\no\in\{2,3,\ldots, 9\}$ for  $\ns\in\{2,3,\ldots, 6\}$. $\RVe{\L}$ for scenarios $(2,5), (2,6),\dots, (2,9)$, $(3,6)$ and $(3,7)$ is very close to 100\% but not exactly 100\%. }
\end{figure}

Quantitative estimate of $\RVe{\L}$ in the 8-dimensional space of $\sP_{\NS}$ for the (2,2) Bell scenario was first determined in Ref.~\cite{Elie_12}. This analysis was then generalized in Ref.~\cite{Duarte_18} to include the (3,2), (4,2), (5,2), (2,3), and the (2,4) Bell scenarios. Among their findings is the observation that for $\no=2$, $\RVe{\L}$ rapidly decreases as $\ns$ increases from 2 to 5. Our findings, as can be seen in \cref{fig:Euclidean:LvsNS}, show that this trend holds also for Bell scenarios with $\no=3,4$ (and possibly $\no=5$). 
\begin{Observation}\label{Obs.LvsNL.fixed.no}
For Bell scenarios with fixed outputs, $\RVe{\L}$ decreases monotonically with increasing $\ns$ (see~\cref{fig:Euclidean:LvsNS}).
\end{Observation}
\noindent Hence, \cref{Obs.LvsNL.fixed.no} generalizes the observation from~\cite{Duarte_18} for $\no=2$. For example,  in the $\no=3$ case, we observe that $\RVe{\L}$ decreases from 93.84\% (for $\ns=2$) to 0.01\% (for $\ns=6$), likewise for the $\no=4$ case, which decreases from 97.41\% (for $\ns=2$) to 6.31\% (for $\ns=5$), etc.

Before discussing this observed trend, note that our estimate for $\RVe{\L}$ in the (2,2) and the (3,2) Bell scenario is consistent with that obtained analytically from the software \lrs~\cite{lrs}, which gives $\RVe{\L}=\frac{16}{17}\approx 94.12\%$ and $\RVe{\L}=\frac{18~176}{29~205}\approx 62.24\%$ respectively. Similarly, our estimate of $\RVe{\L}$ in the $(2,3)$ Bell scenario (see~\cref{Tbl:Results_Detail_Euclidean}) is consistent with that determined from the software \vinci~\cite{Bueler:2000aa}, giving $\RVe{\L}\approx 93.82\%$.

How do we understand the observed decreasing trend? Let us remind that for any Bell scenario ($\ns',\no$) with $\ns'>\ns$, {\em any} {\em sub-correlation} $\vecP$ extracted from $\vecP'$ by considering only $\ns$ out of the $\ns'$ measurement settings (for both Alice and Bob) is a legitimate correlation for the \emph{simpler} Bell scenario ($\ns,\no$). Moreover, for $\vecP'$ to be in $\L$, all these $\binom{\ns'}{\ns}^2$ sub-correlations $\vecP$ extractable from $\vecP'$ must also be Bell-local. 

Let $\RVe{\L}=p$ be the success probability of a Bernoulli trial in the ($\ns,\no$) Bell scenario. If all such $\vecP$ that may be extracted from $\vecP'$ could be thought of as being sampled {\em independently} and {\em uniformly} from the $\NS$ polytope in the ($\ns,\no$) Bell scenario, the success probability of a Bernoulli trial in the ($\ns',\no$) Bell scenario would scale as 
\begin{equation}
	p^{\binom{\ns'}{\ns}^2}= p^{\frac{(\ns'!)^2}{(\ns!)^2[(\ns'-\ns)!]^2}}.
\end{equation}
Applying this na\"ive reasoning to the (2,2) and the (3,2) Bell scenario would suggest a decrease of $\RVe{\L}$ from $\frac{16}{17}\approx 94.12\%$ to $57.95\%$, which is not too far off from our exact finding that $\RVe{\L} =\frac{18~176}{29~205}\approx 62.24\%$ in the (3,2) case. Clearly, part of this discrepancy stems from the fact the sub-correlations $\vecP$  extractable from $\vecP'$ are {\em not} entirely independent from one another -- all these different $\vecP$ share a common input with the other $\vecP$. Moreover, even if all these sub-correlations $\vecP$ are Bell-local, $\vecP'$ may still be Bell-nonlocal.

On the other hand, \cref{fig:Euclidean:LvsNS} shows an opposite trend for Bell scenarios with fixed $\ns$.
\begin{Observation}\label{Obs.LvsNL.fixed.ns} 
	For Bell scenarios with fixed inputs, $\RVe{\L}$ first decreases when $\no$ varies from $2$ to $3$, but increases monotonically thereafter with increasing $\no$ (see \cref{fig:Euclidean:LvsNS}).
\end{Observation}
\noindent This observation generalizes the observation from~\cite{Duarte_18} for $\ns=2$.
Since the way $\RVe{\L}$ changes with increasing $\no$ is opposite to that with increasing $\ns$, it is natural to wonder how $\RVe{\L}$ changes when $\ns = \no = k$ increases. To this end, we have the following observation.
\begin{Observation}\label{Obs.LvsNL.fixed.ns.no}
As $\ns=\no=k$ increases, $\RVe{\L}$ decreases steadily with increasing $k$. That is, the effect of increasing $\ns$ on $\RVe{\L}$ dominates over that of increasing $\no$.
\end{Observation}

\subsection{$\Q$ vs $\NS$ and $\L$}

\subsubsection{Convergence of outer approximations towards $\Qc$}
\label{Sec:Convergence-Q}

Before discussing how $\RVe{\Q}$ changes across different Bell scenarios, let us first make a digression to investigate how well the various $\Q_k$ and $\AQ_\ell$ outer-approximate $\Q$ in {\em each} Bell scenario. Again, their relative volume is a useful figure of merit in this context. From here, we can learn how $\RVe{\T}$  converges to $\Q$ when we consider approximations $\T$ of $\Q$  with increasing complexity. For definiteness, we  make use of the {\em number of real moment variables} involved in the SDP characterization of $\T$ to serve as our measure of complexity. 

\begin{figure*}
  \centering
  \includegraphics[scale=1]{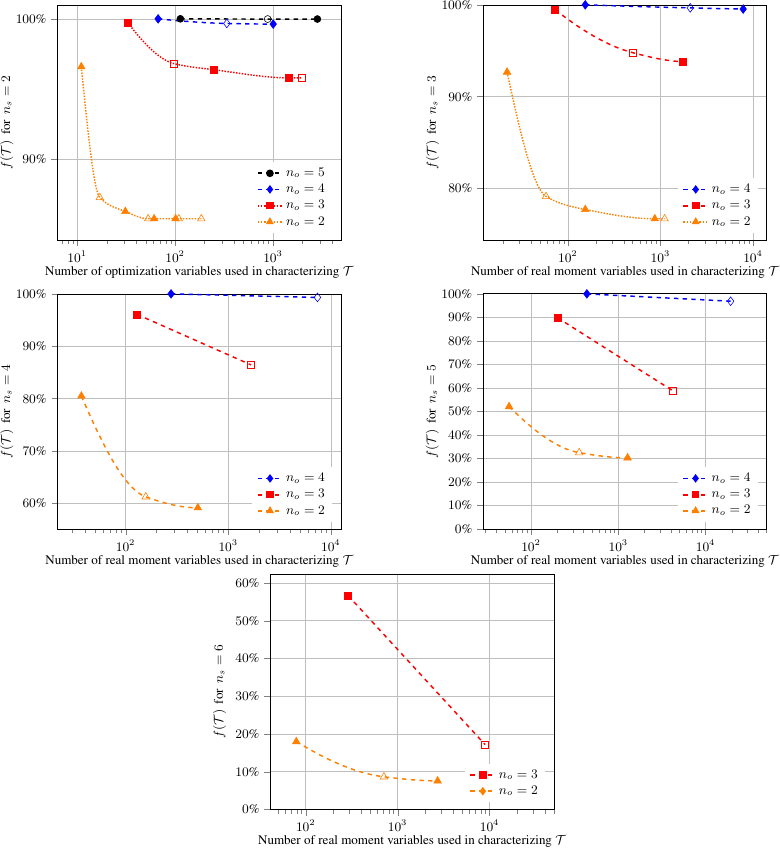}
\caption{\label{fig:Convergent_Euclidean} Numerical estimate of $f(\T)=\frac{\RVe{\T \setminus \L}}{1-\RVe{\L}}$ for various outer approximations $\T$ of  $\Q$, where $\T$ is any member of  $\{\Q_k\}_k$ or $\{\AQ_\ell\}_\ell$. In each plot, we use filled (hollow) markers to represent $\Q_k$ ($\AQ_\ell$). The leftmost filled (hollow) marker corresponds to the first level of the NPA~\cite{NPA} (Moroder {\em et al.}~\cite{Moroder13}) hierarchy $\Q_1$ ($\AQ_1$).
When an increasingly higher level of either hierarchy is considered, the respective SDP characterization involves an increasingly larger number of real (optimization) moment variables (see~\cref{app:Scenarios-Complexity}), thus giving rise to markers that are placed more and more to the right of the plot.  Plots for the (2,6) and (2,7) Bell scenario have been omitted as $f(\T) = 100\%$ for all approximations considered for these Bell scenarios.  As guide for the eye, we have also included a fitting curve (dashed line) in each case. For Bell scenarios with more than three data points, we use the $\mathsf{pchip}$ (piecewise cubic Hermite interpolating polynomial) function in MATLAB to generate the respective fitting curve whereas for the other scenarios, we simply use a straight line to join the two data points.
} 
\end{figure*}

Recall from our discussion in~\cref{Sec:LvsNS} that $\L$ makes up a substantial fraction of $\NS$ for many of the Bell scenarios considered. Thus, to better manifest the convergence graphically, we focus on the nonlocal region of $\NS$, i.e., $\NS\setminus\L$. In other words, for any given outer approximation $\T=\Q_k$ or $\AQ_\ell$, we are interested in the volume of $\T\setminus\L$ {\em relative to} that of $\NS\setminus\L$, i.e., 
\begin{equation}
	f(\T)=\frac{\RVe{\T \setminus \L}}{\RVe{\NS\setminus\L}}=\frac{\RVe{\T \setminus \L}}{1-\RVe{\L}}.
\end{equation}
Evidently, for {\em each} given Bell scenario, there is some $\Q_k$ or $\AQ_\ell$ considered that gives the smallest  $f(\T)$. Denoting them, respectively, by $\Q_*$ and $\AQ_*$, then our tightest approximation to $\Q$ is simply $\Qt:=\argmin_{\T\in\{\Q_*,\AQ_*\}} \RVe{\T}$. The actual $\Qt$ in each case can be read off from the corresponding highlighted entry in~\cref{Tbl:Results_Detail_Euclidean}. We show in~\cref{fig:Convergent_Euclidean} how $f(\T)$ changes  with $\T$ for all those Bell scenarios where we have computed at least two different approximations $\T$'s to $\Q$. For these scenarios, we thus have $\Qt\subsetneq \Q_1$.

\begin{table}[h!]
\begin{center}
    \includegraphics[scale=1]{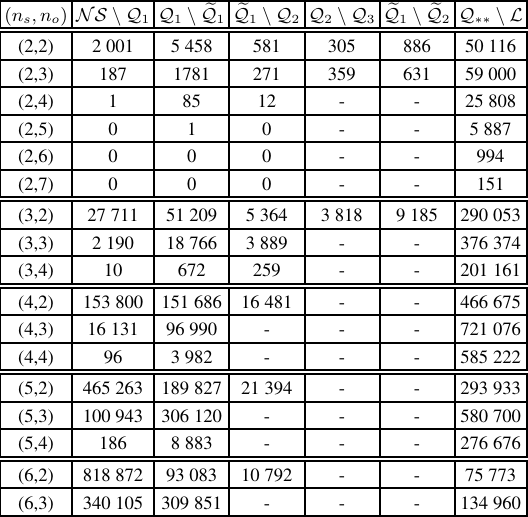}
  \caption{\label{tbl:ConvergenceResults_Euclidean} 
  Summary of the number of correlations $\vecP$ excluded from one approximation of $\Qc$ to a tighter one. The leftmost column gives the Bell scenario. Except for the Bell scenario (5,4) and (6,3), the total number $\Nt$ of correlations sampled from $\NS$ is $10^6$ (see~\cref{Tbl:Results_Detail_Euclidean}). Further to the right, we have, respectively, the number of $\vecP\in\NS$ excluded from $\Q_1$,  the number of $\vecP\in\Q_1$ excluded from $\AQ_1$, the number of $\vecP\in\AQ_1$  excluded from $\Q_2$, the number of $\vecP\in\Q_2$ excluded from $\Q_3$, the number of $\vecP\in\AQ_1$ excluded from $\AQ_2$,  and the number of nonlocal $\vecP$ that lie in our tightest approximation $\Qt$ (see~\cref{Tbl:Results_Detail_Euclidean}). For the Bell scenario (2,2), we have also performed the membership test for $\AQ_3$,  $\Q_4$ and $\AQ_4$. However, no $\vecP$ was further excluded from these higher-level relaxations to $\Qc$. For a graphical representation focussing on the nonlocal region, see~\cref{fig:Convergent_Euclidean}.}
\end{center}
\end{table}

As is evident from the plots (see also~\cref{tbl:ConvergenceResults_Euclidean} and~\cref{Tbl:Results-Difference}), the first level of the NPA hierarchy $\Q_1$ (corresponding to the first {\em filled} symbol on each line) generally does not serve as a very good approximation to $\Q$.  In fact, it largely overestimates $\RVe{\Q}$ for several Bell scenarios.
\begin{Observation}\label{Obs.Q1vsQ}
For two-input Bell scenarios, $\Q_1$ approximates $\Qt$ well but for $\ns>3\ge\no$, $\RVe{\Q_1}$ overestimates $\RVe{\Q}$ by at least 9.69\%. This overestimation even exceeds 30\% in both the (5.3) and (6,4) Bell scenario, see~\cref{Tbl:Results-Difference}.
\end{Observation}

From \cref{fig:Convergent_Euclidean} as well as~\cref{tbl:ConvergenceResults_Euclidean} and~\cref{Tbl:Results-Difference}, we also see that the almost quantum set $\AQ_1$ (corresponding to the first {\em hollow} symbol on each line) generally offers a much better approximation to $\Q$ than $\Q_1$ does. For instance, in all the $\no=2$ scenarios considered, $\Q_1$ occupies between 9 to 20\% more of the nonlocal region compared to $\AQ_1$, whereas in general, $f(\AQ_1)-f(\Qt)< 5\%$. 
Even if we measure according to $\RVe{\T}$, the difference between $\AQ_1$ and $\Qt$ remains small for all the 12 Bell scenarios explored beyond $\AQ_1$.
\begin{Observation}\label{Obs.Q1tildevsQ}
For all the Bell scenarios investigated, $\RVe{\AQ_1}-\RVe{\Qt}$ is larger than 1\% only for the (4,2), (5,2), and the (6,2) Bell scenario. The largest difference is found in the (5,2) Bell scenario, giving  $\approx$ 2.14\%, see~\cref{Tbl:Results-Difference}.
\end{Observation}

\begin{table}[h!]
\begin{center}
    \includegraphics[scale=1]{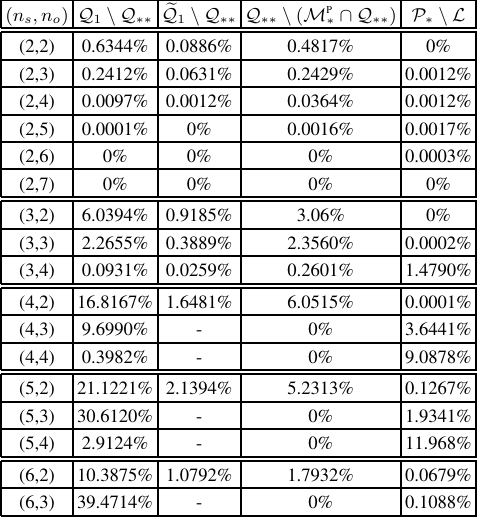}
   \caption{\label{Tbl:Results-Difference} Summary of the difference between the relative volume (RV) of several sets of interest. The leftmost column gives the Bell scenario considered. From the second to the rightmost column, we have, respectively, the difference in the RV between the set associated with the principle of macroscopic locality $\Q_1$ and our tightest approximation to the quantum set $\Qt$, the difference in the RV between the almost quantum set $\AQ_1$ and $\Qt$, the difference in the RV between $\Qt$ and our tightest approximation of the set producible by maximally entangled states with projective measurements $\Mp_*$, and the difference in the RV between our tightest approximation to the set $\P_*$ producible by PPT quantum states and the Bell-local set $\L$. }
\end{center}
\end{table}

\subsubsection{$\Qt$ vs $\NS$}

\begin{figure}[h!]
  \centering
  \includegraphics[scale=1]{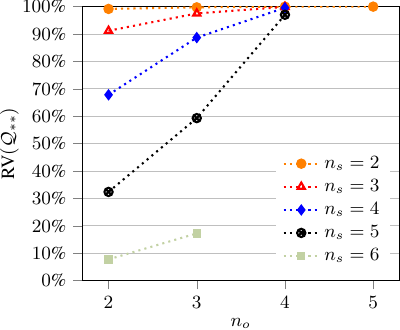}
\caption{\label{fig:Euclidean:QvsNS} Plots of estimated $\RVe{\Qt}$ vs $\no\in\{2,3,\ldots, 5\}$ for  $\ns\in\{2,3,\ldots, 6\}$. The plots  for the (2,6) and (2,7) Bell scenarios are omitted as $\RVe{\Qt} = 100\%$ in both cases.}
\end{figure}

Next, we focus on determining how $\RVe{\Qt}$ varies across the different Bell scenarios.
The trend of how $\RVe{\Qt}$ changes, with $\L$ included, is similar to that of $\L$ shown in~\cref{fig:Euclidean:LvsNS}. 
In a close parallel to \cref{Obs.LvsNL.fixed.no}, we have the following observation for $\Qt$ from~\cref{fig:Euclidean:QvsNS}.
\begin{Observation}\label{Obs.QvsNL.fixed.no}
For Bell scenarios with the same output $\no$, $\RVe{\Qt}$ decreases monotonically with increasing $\ns$.  
\end{Observation}
\noindent Moreover, since $\L$ is a strict subset of $\Q\subset\Qt$, the following observation may have been anticipated from \cref{Obs.LvsNL.fixed.ns}.
\begin{Observation}\label{Obs.QvsNL.fixed.ns}
For Bell scenarios with the same input $\ns$, $\RVe{\Qt}$ increases monotonically with increasing $\no$.  
\end{Observation}

Despite these similarities, there are also  subtle differences. For example, even though for fixed $\ns$, $\RVe{\L}$ generally increases with $\no$, it does so after a dip when $\no$ increases from $2$ to $3$. 
More importantly, for Bell scenarios with $\ns=\no=k$,  while $\RVe{\L}$ appears to decrease monotonically with increasing $k$, $\RVe{\Qt}$ never seems to get far away from 1. 
This suggests that for any given $\ns$, if $\no\ge\ns$ is large enough, a generic $\vecP\in\NS$ is also likely to be a member of $\Q$, i.e., $\RVe{\NS \setminus \Q}$ may become vanishingly small.

\subsubsection{$\Qt$ vs $\L$}

\begin{figure}[h!]
  \centering
  \includegraphics[scale=1]{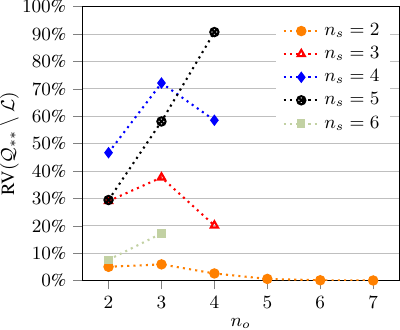}
  \caption{\label{fig:Euclidean:Q-LvsNS} Plots of estimated $\RVe{\Qt \setminus \L}$ vs $\no\in\{2,3,\ldots, 7\}$ for  $\ns\in\{2,3,\ldots, 6\}$. Here $\Qt=\AQ_1$ for the (4,3), (4,4), (5,3), (5,4), and (6,3) Bell scenario whereas $\Qt\subsetneq \AQ_1$ for all the other Bell scenarios considered.}
\end{figure}

What about the the Bell-{\em nonlocal} part of the quantum set, i.e.,  $\Q \setminus \L$? As can be seen from~\cref{fig:Euclidean:Q-LvsNS}, for all $\ns$ investigated, $\RVe{\Qt \setminus \L} = \RVe{\Qt} - \RVe{\L}$ first increases when $\no$ increases from $2$ to $3$. However, for $\ns\le 4$, this difference in RVs decreases for subsequent values of $\no$. Since this is in agreement with the trend of $\RVe{\L}$ shown in~\cref{fig:Euclidean:LvsNS}, the current observation suggests that the trend of $\RVe{\Qt \setminus \L}$ for fixed $\ns\le 4$ is dominated by the trend of $\RVe{\L}$. In contrast, the behavior of $\RVe{\Qt \setminus \L}$ for varying $\ns$ does not seem to follow immediately from that of $\RVe{\Qt}$ nor $\RVe{ \L}$. In particular, for both the $\no=2$ and the $\no=3$ case, we see that $\RVe{\Qt \setminus \L}$ first increases with $\ns$ (from $2$ to $4$) but decreases monotonically after that, which differs from the trend found for the $\no=4$ scenarios. 
\begin{Observation}\label{Obs.QminusL.fixed.no}
$\RVe{\Qt \setminus \L}$ increases monotonically with $\ns$ for the ($\ns$,4) Bell scenarios and reaches $>90\%$ for the (5,4) Bell scenario.  
\end{Observation}
\noindent A large value of $\RVe{\Qt \setminus \L}$ is, {\em a priori}, unexpected as it requires $\RVe{\Qt}$ to be large and $\RVe{\L}$ to be small at the same time. However, both requirements happen to hold for the (5,4) Bell scenario.

Let us further remark that when comparing different Bell scenarios, the value of $\RVe{\Qt \setminus \L}$ need not correlate with the nonlocality of the correlations contained therein. For example, one might expect that the larger $\RVe{\Qt \setminus \L}$, the stronger is the {\em average} resistance of the associated nonlocal correlations to white noise $\vecP_w$. If so, then one might expect the correlations in $\Qt \setminus \L$ for the (5,4) scenario to display the smallest {\em average} white-noise visibility [cf.~\cref{eq:Visibility}] but the results summarized in~\cref{tbl:Euclidean:QvsL} show otherwise.
 
\begin{table}[h!]
\begin{center}
    \includegraphics[scale=1]{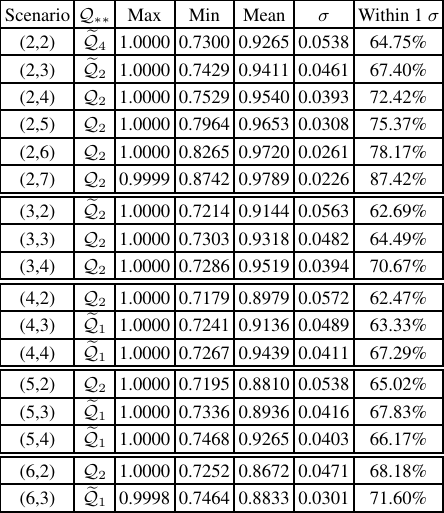}
  \caption{\label{tbl:Euclidean:QvsL} Summary of sample statistics  associated with the distribution of $v^*$ to $\L$ for all those $\vecP$  found to lie in $\Qt\setminus\L$.}
\end{center}
\end{table}
 
An illustration of how various supersets $\T$ of $\L$ contribute towards $\NS \setminus \L$, as measured according to $f(\T\setminus\L)=f(\T)-f(\L)$, can be found in the stacked bar chart displayed in ~\cref{fig:StackedBars_Euclidean} and~\cref{fig:StackedBars_Euclidean2}.

\subsection{Other Naturally Restricted Subsets of $\Q$}

Next, let us consider $\Mp$ and $\P$, two naturally restricted subsets of $\Q$. Again, our tightest outer approximation to these sets are denoted, respectively, by $\Mp_*$ and $\P_*$.

\subsubsection{$\Qt$ vs $\Mp_*$}

Formally, as introduced in~\cref{Sec:Preliminaries}, $\Mp$ is the convex hull of the set of correlations attainable using finite-dimensional maximally entangled states in conjunction with projective measurements. As was first noted in~\cite{Lang_MES}, a hierarchy of (increasingly tighter) outer approximations to $\M$ (the analog of $\Mp$ {\em without} the assumption of measurements being projective) can be obtained via SDPs. In~\cref{app:RelaxationSDP}, we explain our simplified formulation when the measurement are further {\em assumed} to be projective. 

\begin{figure*}
    \centering
    \includegraphics[width=1\linewidth]{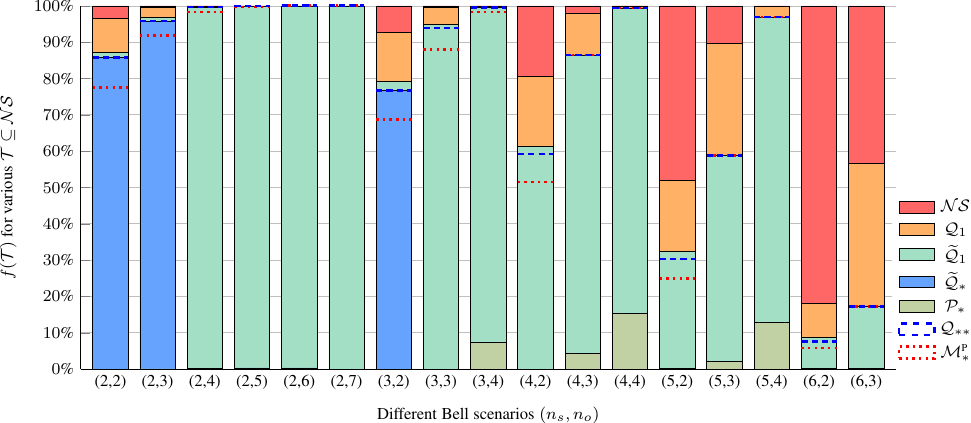}
  \caption{\label{fig:StackedBars_Euclidean}Stacked bar charts showing contributions of various subsets $\T\subseteq\NS$ towards $\T\setminus\L$ where $\T$ is either $\Mp_*$ or any of the subsets listed in the chain of inclusion relations $\P_* \subsetneq \AQ_* \subseteq \AQ_1 \subsetneq \Q_1 \subsetneq \NS$. The RV of a given $\T$ in the nonlocal region, and hence $f(\T)$, is the sum over that due to the other sets contained within it. For example, in the (2,2) Bell scenario, $f(\Q_1)$ corresponds to the orange bar as well as all the other bars (turqoise and blue) stacked below it.  As $\Mp_*$ and $\Qt$ are not directly comparable to some of the aforementioned sets, we use, respectively, dashed and dotted line to represent them separately. Note also that we have only performed computation for higher levels from the hierarchy of Moroder {\em et al.} for the  (2,2), (2,3), and (3,2) Bell scenarios.}
  \vspace{1cm}
    \includegraphics[width=1\linewidth]{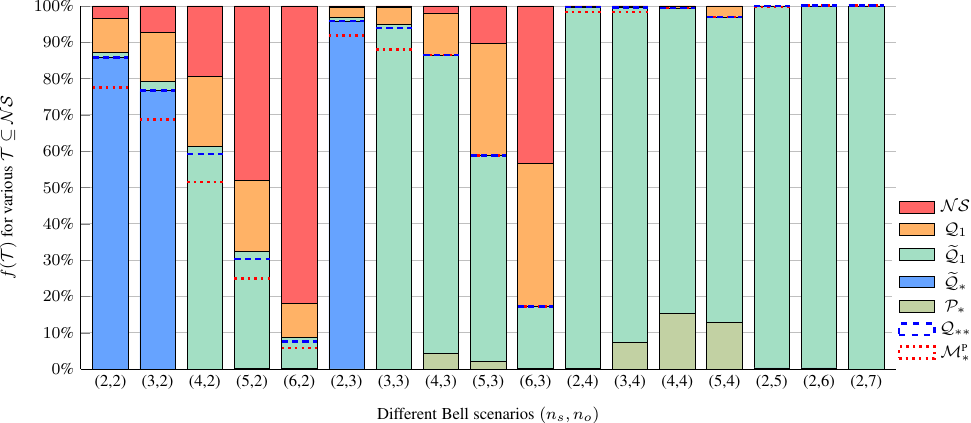}
  \caption{\label{fig:StackedBars_Euclidean2} The stacked bar charts of \cref{fig:StackedBars_Euclidean} sorted first in increasing $\ns$, then followed by increasing $\no$. }  
\end{figure*}

Note that the hierarchy of SDPs used for the computation of $\Qt$ and that of $\Mp_*$ are independent. Consequently, the two sets $\Mp_*$ and $\Qt$ are generally incomparable, i.e., neither of them is necessarily included in the other, despite the fact that $\Mp\subseteq\M\subsetneq\Q$. To quantitatively understand the difference between $\Q$ and $\Mp$, we thus focus on the difference between $\Qt$ and $\Mp_*\cap\Qt$, i.e., the membership test of any given $\vecP$ with respect to $\Mp_*$ is carried out only when it passes the membership test with respect to $\Qt$. For a summary of $\Qt$, $\Mp_*$ involved in the calculation and the sample statistics associated with the distribution of $v^*$ to $\Mp_*$, see~\cref{tbl:Euclidean:QvsMES}.

\begin{table}[h!]
\begin{center}
    \includegraphics[scale=1]{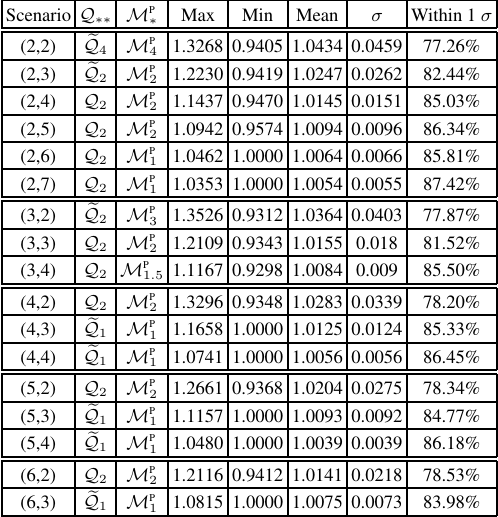}
  \caption{\label{tbl:Euclidean:QvsMES}Summary of sample statistics associated with the distribution of $v^*$ to $\Mp_*$ of all those $\vecP$ found to lie in  $\Qt$. The tightest approximations $\Qt$ and  $\Mp_*$ are included, respectively, in the second and the third column for ease of reference. The notation for $\Mp_h$ is further explained in~\cref{app:Scenarios-Complexity}.}
\end{center}
\end{table}

Our findings (see~\cref{Tbl:Results_Detail_Euclidean} and~\cref{Tbl:Results-Difference}) reveal that the difference between the relative volume of these sets, i.e., $\RVe{\Q_*}-\RVe{\Mp_*\cap \Q_*}$ is rather small. 
\begin{Observation}\label{Obs.QvsMES}
For all the 17 Bell scenarios considered, the largest value of $\RVe{\Q_*}-\RVe{\Mp_*\cap \Q_*}$ ($\approx$ 6.1\%) is found for the Bell scenario (4,2). The Bell scenario (5,2) gives a comparable difference $\approx$ 5.2\% whereas for 12 other Bell scenarios, this difference is less than 0.27\%. In general, the difference decreases with increasing $\no$, see~\cref{Tbl:Results-Difference}.
\end{Observation}

However, if we restrict our attention to only the nonlocal region of $\NS$, then the difference as quantified by $f(\Q_*)- f(\Mp_*\cap \Q_*)$ is most pronounced ($\approx$8.2\%) in the $(2,2)$ Bell scenario, where the original Hardy paradox~\cite{Hardy1993} was proposed, see~\cref{fig:StackedBars_Euclidean}. 

\subsection{$\P_*$ vs $\L$}

Finally, let us focus on the difference between the set of Bell-local correlations $\L$ and that attainable by locally measuring a PPT (entangled) state  $\P$. For a long time, it was believed~\cite{Peres:1999aa} that no bound entangled state can violate a Bell inequality. Since all PPT entangled states are bound entangled~\cite{Horodecki1998}, this so-called Peres conjecture would imply that $\P=\L$. Indeed, in the simplest (2,2) Bell scenario, it was shown by Werner and Wolf~\cite{PPT_CHSH} that these sets do coincide. Even for the (3,2) Bell scenario, numerical results from~\cite{Moroder13} again indicate that $\P=\L$.

That this conjecture does not hold in full generality was first shown~\cite{Vertesi:PRL:2012} in a tripartite Bell scenario using a three-qubit bound entangled state. Later, the conjecture was also disproved~\cite{Vertesi_PPTState} in a {\em bipartite} setting by considering a two-qutrit PPT entangled state and an asymmetric Bell scenario $\{[2~2~2]~[3~2]\}$, where the number of entries in each square bracket denotes the number of settings for each party and the actual numbers listed are the number of measurement outcomes for each setting.\footnote{The notation adopted here for this asymmetric Bell scenario follows that introduced in Ref.~\cite{Barnea2013}.} Consequently, for any symmetric Bell scenario with $\ns,\no\ge 3$, we must have  $\L\subsetneq\P$. Moreover, it can be shown from the results of Ref.~\cite{Vertesi_PPTState} and~\cref{eq:Visibility} that the corresponding $\vecP\in\P$ has a visibility $v^*$ to $\L$ that is approximately 0.9996. 

Here, we make use of the outer approximations proposed in~\cite{Moroder13} to quantitatively survey the difference between $\P$ and $\L$.
Interestingly, even though $\P_*$ only outer approximates $\P$, we see from~\cref{Tbl:Visibility:PtoL} that for the (2,2) and the (3,2) Bell scenario, these approximations work extremely well: among all the $10^6$ samples generated for each of these Bell scenarios, there is not even a single $\vecP\not\in\P_*$ that lies outside $\L$. This is, of course, consistent with the known results given, respectively, in Ref.~\cite{PPT_CHSH} and Ref.~\cite{Moroder13}. In fact, for the (2,2) scenario, even the lowest-level approximation given by $\P_1$ does not give rise to any $\vecP\not\in\L$ from $10^6$ samples.

In contrast, except the (2,7) Bell scenario, we do find instances of $\vecP\in\P_*\setminus\L$  for {\em all} {\em other} Bell scenarios investigated, see~\cref{Tbl:Visibility:PtoL}. 
For this exceptional case, we note from~\cref{Tbl:Results_Detail_Euclidean} that $\L$ almost spans the entire non-signaling set, thus leaving very little room for $\NS\setminus\L$, let alone $\P\setminus\L$. For most of the other Bell scenarios, 
even though we know from~\cref{Tbl:Visibility:PtoL} that $\P_*\neq \L$, the difference between $\RVe{\P_*}$ and $\RVe{\L}$ is  tiny, if not vanishingly small.

\begin{Observation}\label{Obs.PvsL}
For all the 17 Bell scenarios considered, the largest value of $\RVe{\P_*\setminus\L}$ ($\approx$ 12\%) is found for the Bell scenario (5,4). The Bell scenario (4,4) shows a comparable difference of $\approx$ 9.1\% whereas for 12  other Bell scenarios, this difference is less than 0.15\%. Except the $\ns=2$ Bell scenarios, $\RVe{\P_*\setminus\L}$ increases with increasing $\no$, see~\cref{Tbl:Results-Difference}.
\end{Observation}

Zooming into the behavior of the individual correlation, we see  from~\cref{Tbl:Visibility:PtoL} that within these approximations, one can find $\vecP\in(\P_*\setminus\L)$ that are far more robust with respect to the mixing with white noise in more complex Bell scenarios. For example, in the (4,4) Bell scenario, the most robust $\vecP\in\P_*$ was found to give a visibility of 0.9037, as compared with a visibility of 0.9996 found~\cite{Vertesi_PPTState} in the (3,3) Bell scenario.

\begin{table}[h!]
\begin{center}
    \includegraphics[scale=1]{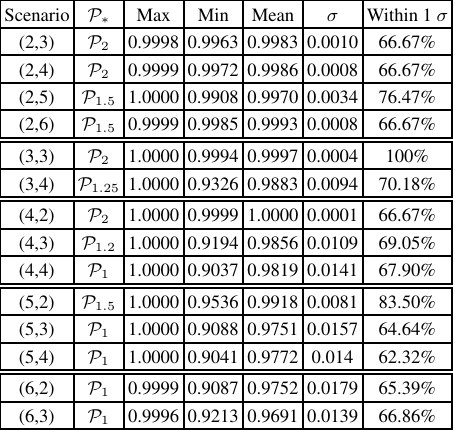}
  \caption{\label{Tbl:Visibility:PtoL} 
  Summary of sample statistics associated with the distribution of $v^*$ to $\L$ of all those $\vecP$ found to lie in  $\P_*\setminus\L$. For the Bell scenarios (2,2), (2,7) and (3,2), we do not find any correlation $\vecP\in\P_* \setminus \L$,  thus the respective rows are omitted. The tightest approximation $\P_*$ is included in the second column for ease of reference. The notation for $\P_z$ is further explained in~\cref{app:Scenarios-Complexity}.}
\end{center}
\end{table}

\section{Discussions}
\label{sec:Discussion}

In the studies of Bell-nonlocality~\cite{Brunner_RevModPhys_2014} and its applications in device-independent quantum information~\cite{Scarani_DIQI_12} (DIQI), one often exploits the geometrical features associated with the correlation $\vecP$ or the set(s) in which they belong to draw conclusions. For example, the convexity of $\L$, $\Q$, and $\NS$ and the possibility to describe $\L$ and $\NS$ using a finite number of extreme points are often invoked to simplify the analyses in DIQI. Nonetheless, despite the numerous efforts devoted to these lines of research, much about the geometry of these sets remain to be understood, see, e.g., Ref.~\cite{Goh2018}. 

In fact, even though the inclusion relations $\L \subsetneq \Mp \subsetneq\Q \subsetneq \AQ_{1}\subsetneq \Q_1\subsetneq \NS$ and $\L\subseteq \P\subsetneq \Q$ are long known, we do not have any quantitative understanding of their difference beyond the extremely limited exploration carried out in Refs.~\cite{Cabello_05,Elie_12,Duarte_18} for $\L$, $\Q$, and $\NS$. In this work, by determining the relative volume (RV) of these sets or their outer approximations, we aim to fill this gap by putting some of the intuitions that the community has developed over the years on more quantitative grounds. As $\RVe{\T}$ can be interpreted as the success probability of a Bernoulli trial, if $\RVe{\T}\ge 100\%-\epsilon$ for some $\T\subseteq\NS$, the operational task of finding a correlation in $\NS\setminus\T$ by a uniform sampling in $\NS$ can only succeed with probability at most $\epsilon$. En route to the above goal, several intriguing observations are noted.

Firstly, our results suggest that, the fraction of $\L$ in $\NS$ becomes vanishingly small (\cref{Obs.LvsNL.fixed.no}) as we {\em fix} the number of outputs $\no$ but {\em increases} the number of inputs $\ns$. A similar monotonic behavior is also observed for the macroscopically-local set $\Q_1$, the almost-quantum set $\AQ_1$ (see~\cref{Tbl:Results_Detail_Euclidean}), as well as our tightest approximation to the quantum set $\Qt$ (\cref{Obs.QvsNL.fixed.no}). Although the two-output case of \cref{Obs.LvsNL.fixed.no} was already noted in Ref.~\cite{Duarte_18}, our results provide further evidence this trend could well be generic, i.e., independent of the actual value of $\no$. Moreover, we demonstrate how a combinatoric argument can be used to understand this decreasing trend.

Interestingly, apart from a dip at the beginning, an essentially opposite trend for $\RVe{\L}$ is found when we {\em fix} the number of inputs $\ns$ but {\em increases} the number of outputs $\no$ (\cref{Obs.LvsNL.fixed.ns}). This increasing trend is even found to be {\em monotonic} for several supersets of $\L$, including $\Q_1$, $\AQ_1$ (see~\cref{Tbl:Results_Detail_Euclidean}), and $\Qt$ (\cref{Obs.QvsNL.fixed.ns}). Again, this suggests that the trend---already noted in Ref.~\cite{Duarte_18} for the special case of $\ns=2$---could well be generic and independent of the specific value of $\ns$. If so, then for any given $\ns$, the quantum set $\Q$ and its superset $\NS$ may become essentially indistinguishable when $\no$ is sufficiently large. While we do not have an intuitive explanation for this observation, it seems plausible that it can lead to interesting consequences on the power of nonlocal quantum resources.

What about the quality of various outer approximations of $\Q$, which is especially relevant for a variety of tasks in DIQI? Our results suggest that for two-input Bell scenarios, $\Q_1$ already provides a  superb outer approximation, with $\RVe{\Q_1}-\RVe{\Qt}<0.7\%$, but for Bell scenarios with $\ns>3\ge \no$, the reliability of $\Q_1$ as an outer approximation becomes questionable (\cref{Obs.Q1vsQ}). In these cases, the difference $\RVe{\Q_1}-\RVe{\Qt}$ can even get as large as $\approx 39.5\%$. In contrast, the almost-quantum set $\AQ_1$, as its name suggests, gives consistently a tiny deviations, if at all, from our tightest quantum approximation (\cref{Obs.Q1tildevsQ}). Still, our results (see~\cref{Tbl:Results-Difference}) suggest that a more noticeable difference may be found in a Bell scenario with two outputs but a larger number of inputs, say, $\ns=5$.

Moving on to naturally restricted subsets of $\Q$, it is known from the work of Ref.~\cite{Liang2011} and Ref.~\cite{Vidick2011} that $\Mp\subseteq\M\subsetneq\Q$, respectively, in the (2,2) and the (3,2) Bell scenario. Our results indicate that for fixed $\ns$, the difference between $\Mp$ and $\Q$ may diminish following the increase in $\no$ (\cref{Obs.QvsMES}). If so, in a Bell scenario where $\ns\ll\no$, it may be sufficient to consider only finite-dimensional maximally entangled states in conjunction with projective measurements for various tasks in DIQI. However, our observation should not be taken to imply that $\Mp\to\Q$ for {\em any} Bell scenario with large enough $\no$, see, e.g., Ref.~\cite{Junge:2011aa}.

In contrast, our results do suggest that the difference between $\P$ and $\L$ may become noticeable only when the Bell scenario involved is sufficiently complex (\cref{Obs.PvsL}). In particular, for the $(\ns,2)$ and $(2,\no)$ Bell scenarios that we have investigated, the respective $\RVe{\P_*}-\RVe{\L}$  is always found to be tiny  ($<0.17\%$). For all but two of the remaining Bell scenarios, this difference  is also never more than $3.7\%$. That leaves us only with the  (4,4) and the (5,4) Bell scenario---among all those computed---as the most promising candidate for an experimental demonstration of the Peres conjecture violation. Still, even though we get, respectively, $\RVe{\P_*\setminus\L}\approx 9\%$ and $\approx 12\%$, and a minimum visibility of $\approx 0.90$ for both Bell scenarios, further investigation is clearly needed to confirm its experimental  viability.

Apart from this, a few other closely-related research directions may be worth pursuing. For example, even though both $\P$ and $\Mp$ are known to be subsets of $\Q$, their precise relationship is not known. Intuitively, one would expect $\P$ to be a strict subset of $\Mp$, which is supported by our observation that $\RVe{\Qt}-\RVe{\Mp_*}$  and $\RVe{\P_*}-\RVe{\L}$ are typically small, but proving this does not seem to be trivial. Evidently, a comprehensive estimation of the relative volume of various naturally restricted subsets in the multipartite setting is also desirable. Due to the rich structure in multipartite entanglement~\cite{He:PRX:2018} and multipartite nonlocality~\cite{Curchod2015}, there will be many more natural subsets of $\NS$ to consider in those cases.

Another direction that deserves further investigation is that related to the choice of metric in our sampling. In this work, we have opted for the Euclidean metric defined in the space of $\sP_{\NS}$. However, since the space of interest $\sP$ is a probability space, an arguably more natural~\cite{Acin:Private} metric is the so-called Fisher (information) metric $\dSf^2$. For a set of {\em unconditional} probability distributions $q_i$ such that $q_i\ge 0$ and $\sum_i q_i =1$, the Fisher metric and hence the corresponding volume element are given, respectively, by~\cite{Acin_Fishermetric}:
\begin{equation}\label{eq:FisherMetric}
    \dSf^2 = \sum_i \frac{\dd q_i^2}{q_i} \quad \text{ and }\quad
    \dVf = \prod_i\left( \frac{\dd q_i}{\sqrt{q_i}}\right). 
\end{equation}
Recall from~\cref{Sec:Preliminaries} that we are interested in {\em conditional} distributions lying in the non-signaling subspace, cf.~\cref{eq:NS}. These requirements, together with a nontrivial dependence of the metric on the coordinate in $\sP$, however, make it unclear how we can perform a uniform sampling according to this metric (e.g., using an existing software package like \cprnd). Similarly, a careful reader would have noticed that if we work in the space of $\sP$, the uniform distribution $\vecP_w$ has the same Euclidean distance to all extreme points of $\L$ (see~\cref{app:DistanceOfExtremalPoints}) but if we make use of the parametrization given in~\cref{Eq:CG}, then this invariance is lost. As such, it would be interesting to find a parameterization of $\sP_{\NS}$ where this invariance is preserved and repeat the calculation performed here.

Finally, recall that in our studies of convergence of the outer approximations of $\Q$ (see~\cref{Sec:Convergence-Q}), we use the hierarchies of SDPs defined by NPA~\cite{NPA} as well as those given in Moroder {\em et al.}~\cite{Moroder13}. These are, however, not the only outer approximations of $\Q$ that discussed in the literature. For example, the SDPs defined by Berta {\em et al.}~\cite{Berta:2016aa} are also known to define a converging hierarchy of outer approximations. Moreover, in comparison with the NPA hierarchy, the approximations of Berta {\em et al.} are known to be (possibly) tighter as they include further non-negativity requirement of certain elements of the moment matrix (see also Ref.~\cite{Sikora:2017aa}). It could thus also be interesting to investigate, how this, and other hierarchies of outer approximations (e.g., those discussed in~\cite{SLChen18}) appear to converge.

\acknowledgements
  We are grateful to Tulja Varun Kondra for his contribution at the initial stage of this research project, to Antonio Ac\'{\i}n, Jean-Daniel Bancal, Elie Wolfe, and Denis Rosset for their suggestions on how uniform sampling on convex sets (of correlations) may be performed. Part of this work was completed during PSL's visit to the Institute of Photonic Sciences (ICFO), Spain and YCL's visit to the Perimeter Institute for Theoretical Physics, Canada. The hospitality of both institutes is greatly appreciated. This work is supported by the Ministry of Science and Technology, Taiwan (Grants No. 104-2112-M-006-021-MY3, 107-2112-M-006-005-MY2, and 109-2112-M006-010-MY3), the EU (QuantERA eDICT) and the National Research, Development and Innovation Office NKFIH (No. 2019-2.1.7-ERA-NET-2020-00003).


\appendix

  
\section{Approximation to the $\Mp$}
\label{app:RelaxationSDP}

In this section, we explain how our approximations to the {\em convex hull} of the set of correlations attainable by finite-dimensional maximally entangled quantum states $\M$ are defined. As was pointed out in Ref.~\cite{Lang_MES}, $\M$ can be {\em outer} approximated by a hierarchy of correlations, each amenable to an SDP characterization. Here, we focus on a subset $\Mp$ of $\M$, where the local POVM elements are further assumed to projectors. 

To appreciate how the hierarchy works, let us first remind that for any local POVM element $E_{a|x}$ $E_{b|y}$ acting on a bipartite $d$-dimensional maximally entangled state $\ket{\Psi_d} = \frac{1}{\sqrt{d}}\sum_{i=1}^d\ket{i}\ket{i}$, 
we have:
\begin{equation}\label{eq:maxPabxy}
\begin{split}
  P(a,b|x,y) = \bra{\Psi_d} E_{a|x}\otimes E_{b|y}\ket{\Psi_d}
  = \tfrac{{\rm tr} \left(E_{a|x}E_{b|y}\tp\right)}{d},
\end{split}
\end{equation}
where $(\cdot)\tp$ denotes transposition.

The essence of the characterization of $\M$, and hence of our characterization of $\Mp$, is an approximation of the trace function in~\cref{eq:maxPabxy} by a linear function acting on the POVM elements. To this end, let us define $M_0=\{\Id\}$, which is the set consisting of only the $d$-dimensional identity operator, and its union with a set of {\em projective} POVM elements 
\begin{equation}
	M_1=\{\Id\}\cup \{\tilde{E}_{a|x}\}_{a,x} \cup \{\tilde{E}_{b|y}\}_{b,y}
\end{equation}
where $a,b=1,2,\ldots,\no-1$, $x,y=1,2,\ldots,\ns$. Here, the projective nature of the POVM elements implies 
\begin{equation}
	\tilde{E}_{a|x}\tilde{E}_{a'|x}=\tilde{E}_{a|x}\delta_{a,a'},\quad
	\tilde{E}_{b|y}\tilde{E}_{b'|y}=\tilde{E}_{b|y}\delta_{b,b'}.
\end{equation}
 More generally, for any positive integer $k> 1$, let us {\em define} the set of operators with degree $k$ or less as
\begin{equation}
	M_k = \cup_i\, \left\{\Pi_i(\tilde{E}^1\dots \tilde{E}^k)\right\},
\end{equation}
where  the union is over all possible permutations $\Pi_i$ of $k$-fold product of operators chosen from $M_1$. Notice that as both $\tilde{E}_{a|x}$ and $\tilde{E}_{b|y}$ act on the same Hilbert space, they generally do not commute.
 
Now, in analogy to the work of Ref.~\cite{Lang_MES}, we {\em define} a bipartite correlation $\vecP$ to be a member of $\Mp_k$, $k\ge 1$, if there exists an integer $d\ge 2$ and a linear functional $L: M_k\times M_k \rightarrow \sP_{\NS}$ such that the following properties hold:
\begin{enumerate}
\item $L(\Id) = 1$.
\item $L(ff^\dagger) \ge 0$ for any $f \in M_k$.
\item $L(f \tilde{E}_{a|x} f^\dagger), L(f \tilde{E}_{b|y} f^\dagger) \ge 0$ for any $f \in {M}_{k-1}$.
\item $L(f \tilde{E}_{a|x} f^\dagger \tilde{E}_{a'|x'}), L(f \tilde{E}_{b|y} f^\dagger \tilde{E}_{b'|y'})\ge 0$ and $\\ L(f \tilde{E}_{a|x} f^\dagger \tilde{E}_{b|y}), L(f \tilde{E}_{b|y} f^\dagger \tilde{E}_{a|x})\ge 0$ for any $f \in {M}_{k-1}$.
\item $L(ST) = L(TS)$ where $ST\in M_{2k}$. 
\item $L(\tilde{E}_{a|x}\tilde{E}_{b|y}) = P(a,b|x,y)$ for all $a,b,x,y$.
\end{enumerate}

Clearly, if $\vecP\in\Mp$, then by~\cref{eq:maxPabxy}, a linear functional satisfying all the above properties is guaranteed to exist by taking $L(\cdot)=\frac{1}{d}\tr(\cdot)$  and setting $\tilde{E}_{a|x} = E_{a|x}$ and $\tilde{E}_{b|y} = E\tp_{b|y}$. In other words, $\vecP\in\Mp\implies \vecP\in\Mp_k$ for all $k\ge 1$. Importantly, for any given integer $k\ge 1$, the membership of any given $\vecP\stackrel{?}{\in}\Mp_k$ can be determined by solving an SDP that amounts to requiring  the existence of a positive semidefinite moment matrix $\Gamma$ with its entries given by $\Gamma_{ij} = L(f_if_j^\dagger)$ where $f_i, f_j\in M_k$ and where all these entries are required to satisfy the linear constraints listed above. 

A few remarks are now in order. Firstly, as oppose to the NPA hierarchy where the convexity of $\Qc$ (and hence of $\Q_k$) is promised by {\em not} restricting the underlying Hilbert space dimension, convexity has to be assumed in the formulation of $\M$ or $\Mp$ by considering the convex hull of the set of correlations attainable from $\ket{\Psi_d}$. Secondly, in the formulation of $\M$, the projective nature of the POVM elements cannot be taken for granted, since a na\"ive application of Naimark's extension~\cite{Peres:1990aa} does not guarantee that the state to which the extended projective POVM elements are applied is maximally entangled. 

Consequently, our characterization differs from that given in Ref.~\cite{Lang_MES} in two aspects: (1) our formulation assumes that $\tilde{E}_{a|x}$ and $\tilde{E}_{b|y}$ are projective while that of Ref.~\cite{Lang_MES} does not (2) the formulation given in Ref.~\cite{Lang_MES} actually imposes in property 3. above the more stringent requirement that $f$ can be {\em any} linear combination of elements in $M_{k-1}$. Since one of these differences is more constraining while the other is less constraining, our hierarchy $\Mp_k$ is neither a subset nor a superset of the corresponding set $\Q^k_+$ defined in Ref.~\cite{Lang_MES}. Empirically, we have also found that if we keep difference (2) and drop the assumption of $\tilde{E}_{a|x}$, $\tilde{E}_{b|y}$ being projective, then the resulting relaxation of $\Q^k_+$ appears to be hardly constraining. Finally, notice that except for some additional positivity requirement due to property 4 above, the SDP for $\Mp_1$ is the same as that for NPA level 1.

\section{Sampling methods} \label{app:SamplingMethod}

To obtain uniformly sampled correlations in $\NS$ using the \cprnd~function, we make use of the minimal parametrization given in~\cref{Eq:CG}. For  this purpose, it suffices to input to  \cprnd~(1) the total number of samples $\Nt$ required (2) the sampling algorithm to be employed,\footnote{In our work, we use ``Gibbs" sampler that gives, empirically, better convergence properties than the default ``hit-and-run" sampler.} and (3) a complete description of the nonsignaling polytope $\NS$ in terms of its positivity (facet) constraints.

Explicitly, these positivity constraints are
\begin{equation}
  \label{eq:ConstraintsEuclidean_trivial}
    P(a|x)  \ge 0,\quad
    P(b|y)  \ge 0,\quad
    P(a,b|x,y) \ge 0, 
  \end{equation}
in addition to
\begin{equation}
  \label{eq:ConstraintsEuclidean1}
    P(a=\no|x) = 1 - \sum_{a'=1}^{\no-1}P(a'|x) \ge 0,\quad
    P(b=\no|y) = 1 - \sum_{b'=1}^{\no-1}P(b'|y) \ge 0,
\end{equation}
and
\begin{equation}
  \label{eq:ConstraintsEuclidean2}
  \begin{aligned}
    P(a,b=\no|x,y) &= P(a|x) - \sum_{b'=1}^{\no-1}P(a,b'|x,y) \ge 0,\\
    P(a=\no,b|x,y) &= P(b|y) - \sum_{a'=1}^{\no-1}P(a',b|x,y) \ge 0, \\
    P(a=\no,b=\no|x,y)& = 1- \sum_{a'=1}^{\no-1}P(a'|x) - \sum_{b'=1}^{\no-1}P(b'|y) 
     + \sum_{a',b'=1}^{\no-1}P(a',b'|x,y) \ge 0 
    \end{aligned}
\end{equation}
for all $a,b\in\{1,\ldots, \no-1\}$ and $x,y\in\{1,2,\ldots, \ns\}$.

\section{Bell scenarios considered and the complexity involved in the characterization of various sets}
\label{app:Scenarios-Complexity}

\begin{table*}[t!]
\begin{center}
    \includegraphics[width=1\linewidth]{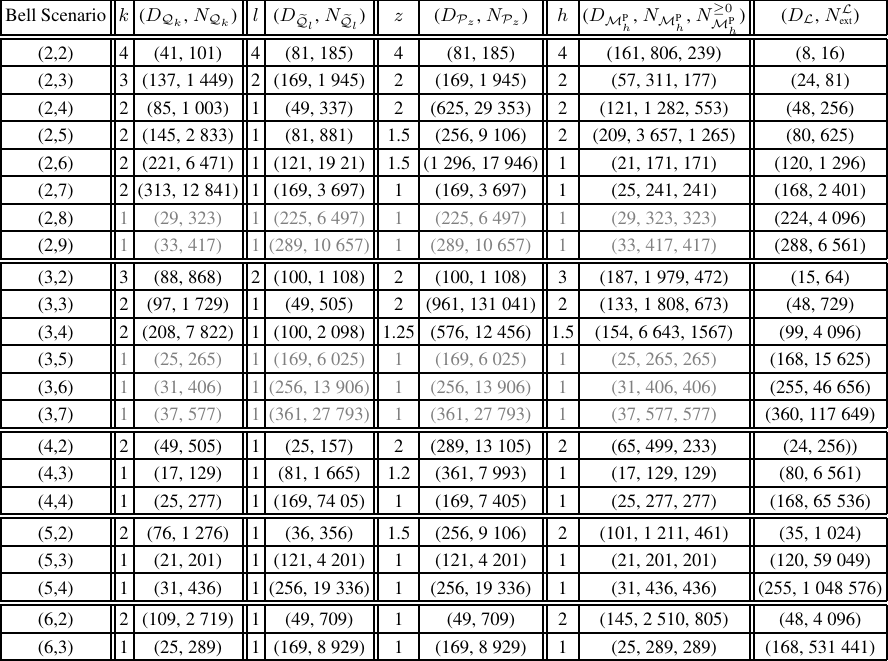}
  \caption{\label{Tbl:Data} Summary of the Bell scenarios considered in this work (leftmost column), the highest level of the SDP relaxation considered (second, fourth, sixth, and the eighth column), and the corresponding parameters characterizing the complexity of the computation (third, fifth, seventh, and the ninth column). In particular, for the NPA hierarchy (the second and third column), the Moroder hierarchy for $\Q$ (the fourth and fifth column), the Moroder hierarchy for $\P$ (the eighth and ninth column), we provide in bracket the size of the respective moment matrix $D$ and the number of (real) moment variables involved in the corresponding optimization.  Similarly, for the hierarchy of SDPs characterizing $\Mp$ (the fifth and sixth column), we provide in bracket the size of the respective moment matrix $D_{\T}$, the number of (real) moment variables $N_\T$ involved in the corresponding optimization, and the number of moments that are further required to be non-negative $N_{\Mp_h}^{\ge 0}$. In the last column, we list, accordingly, the key parameters characterizing the complexity of the linear program involved in solving the membership problem $\vecP\stackrel{?}{\in}\L$, i.e., the size $D_{\L}$ of $\vecP$, which equals to the dimension $d$ of $\NS$ and the number of extreme points $N_\text{\tiny ext}^{\L}$ of the respective local polytope $\L$. For completeness, we also include in gray the parameters characterizing the complexity of the various level-1 SDPs for the (2,8), (2,9), (3,5), (3,6), (3,7) Bell scenarios that we did not compute.}    
\end{center}  
\end{table*}

We consider bipartite Bell scenarios where each party has $\ns$ measurement settings and where each measurement gives $\no$ outcomes. Bearing in mind that the generating POVM elements may be chosen to be projectors, the size of the Moroder~\cite{Moroder13} level $\ell$ moment matrix, denoted by $\Dml$, can be shown to be:
\begin{equation}
  \Dml=\left[1+ \sum_{j=1}^\ell \ns(\ns-1)^{j-1}(\no-1)^j\right]^2.
\end{equation}
Similarly, the size of the NPA level $k$ moment matrix, denoted by $\Dnpal$ can be shown to be:
\begin{equation}
  \Dnpal = 1 + 2\ns\sum_{j=1}^k (\ns-1)^{j-1}(\no-1)^j 
  +\ns^2\sum_{j=2}^k (j-1)(\ns-1)^{j-2}(\no-1)^j,
\end{equation}
where the first sum consists of only contributions of $k$-fold products of operators from the same party, the last sum consists of $k$-fold products of operators originating from both parties, and the factor $(j-1)$ in the last sum accounts for different possibilities in terms of the number of Alice's and Bob's operators.

Finally, the size of the moment matrix corresponding to the characterization of level $h$ of $\Mp$, denoted by $\Dmesl$ is:
\begin{equation}
  \Dmesl = 1 +2\ns\sum_{j=1}^h (2\ns-1)^{j-1}(\no-1)^j.
\end{equation}
From the expressions given above, it is clear that for all these hierarchies, the size of the SDP moment matrix increases exponentially with the level of the hierarchy. As such, due to limitation in computational resources, it is also expedient to consider intermediate, non-integer level of these hierarchies in order to obtain a tighter approximation. 

For both hierarchies of SDPs due to Moroder {\em et al.}, let  
\begin{equation}
	D_{\text{\scriptsize local},\ell}=1+ \sum_{j=1}^\ell \ns(\ns-1)^{j-1}(\no-1)^j.
\end{equation}
Furthermore, let us denote by $\lfloor q \rfloor$ be the integer part of a positive number $q\ge 1$. Then, we say that an outer approximation is of level $q$ if the considered moment matrix $\Gamma_q$ contains $\Gamma_{\lfloor q \rfloor}$ as a submatrix while $\Gamma_q$ itself is a submatrix of $\Gamma_{\lfloor q \rfloor+1}$. Moreover, the moment matrix of $\Gamma_q$ is formed by considering only (approximately) the first $(q - \lfloor q \rfloor)(D_{\text{\scriptsize local}, \lfloor q \rfloor+1}-D_{\text{\scriptsize local}, \lfloor q \rfloor})$ level-$(\lfloor q \rfloor+1)$ local operators in addition to {\em all} the level-$\lfloor q \rfloor$ local operators.\footnote{Note that our level-$\ell$ local operators $A_{a_1|x_1}A_{a_2|x_2}\cdots A_{a_\ell|x_\ell}$ are ordered by first increasing the index of $a_1$, followed by $x_1$, followed by $a_2$, etc.}

In a similar manner, for the hierarchy defined in~\cref{app:RelaxationSDP}, we say that an outer approximation for $\Mp$ is of an intermediate level $q$ if the considered moment matrix $\Gamma_q$ contains $\Gamma_{\lfloor q \rfloor}$ as a submatrix while $\Gamma_q$ itself is a submatrix of $\Gamma_{\lfloor q \rfloor+1}$. Moreover, the moment matrix of $\Gamma_q$ is formed by taking the upper-left submatrix of $\Gamma_{\lfloor q \rfloor+1}$ with (approximately) $D_{\Mp_{\lfloor q \rfloor}}+(q - \lfloor q \rfloor)(D_{\Mp_{\lfloor q \rfloor+1}}-D_{\Mp_{\lfloor q \rfloor}})$ rows and columns.\footnote{Here, our level-$h$ operators $A_{a_1|x_1}\cdots A_{a_{h-1}|x_{h-1}}A_{a_h|x_h}$ are ordered by first increasing the index of $a_{h}$, followed by $x_{h}$, followed by $a_{h-1}$, etc. Moreover, we adopt the convention that  $A_{a_i|x_i}=B_{a_i|x_i-\ns}$ for $x_i>\ns$ where $A_{a|x}\leftrightarrow \tilde{E}_{a|x}$,  $B_{b|y}\leftrightarrow \tilde{E}_{b|y}$, see~\cref{app:RelaxationSDP}.}

Of course, the size of the moment matrix $D_\T$ for a set $\T$ is not only the parameter that determines the computational resource required to solve each of these SDPs. In particular, for the membership test corresponding to  $T\in\{\Q_k, \AQ_\ell, \Mp_h, \P_z\}$, the number $N_\T$ of real variables (independent moments) involved in the corresponding moment matrix also plays a crucial role. In the case of $\Mp_h$, the number $N_{\Mp_h}^{\ge 0}$ of moments that are further required to be non-negative also play a part in the complexity of the problem. In~\cref{Tbl:Data}, we provide a summary of the Bell scenarios considered in this work as well as these key parameters relevant to solving the corresponding optimization problems.

\begin{figure*}
  \centering
  \includegraphics[scale=1]{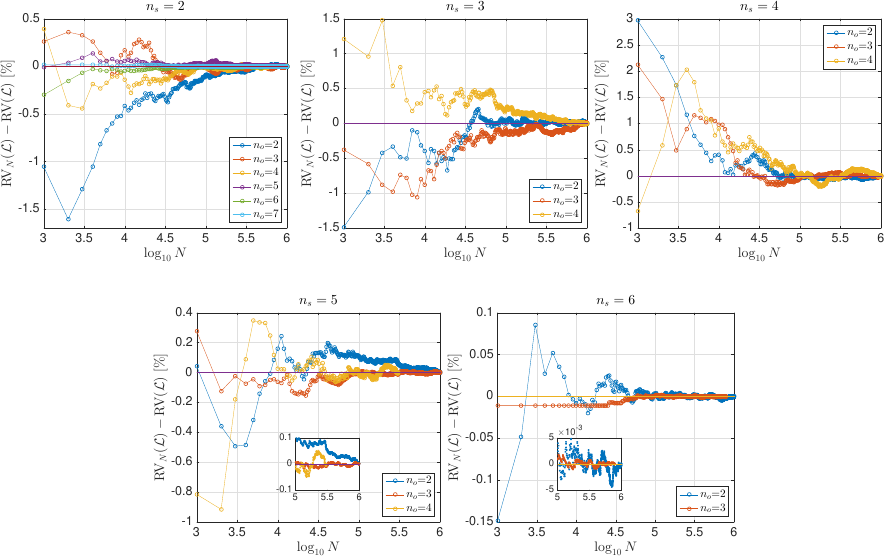}
  \caption{\label{fig.Conv.ns.sample}  Plots showing how the {\em estimated} relative volume $\RVe{\L}$ converges to the reported value as the number of samples $N$ used in the estimation increases towards $\Nt$. More precisely, in each subfigure, we plot  RV$_N(\L)-\RVe{\L}$ as a function of $\log_{10}N$, where RV$_N(\L)$ is the value of $\RVe{\L}$ estimated using the first $N$ correlations $\vecP$ sampled uniformly from $\NS$. As a reference, note that $\RVe{\L}$ lies, respectively, within $[93\%,~ 100\%]$, $[59\%,~ 100\%]$, $[16\%,~ 81\%]$, $[1\%,~ 7\%]$, and $[0.01\%,~ 0.15\%]$ for $\ns=2,3,4,5,$ and $6$.
For these two latter cases, an inset showing the plots from $\log_{10}N\in[5,~6]$ is included to show that $\RVe{\L}$ has converged to within $10\%$ of its value.  As an alternative way to see that our estimated RVs have converged well, one can also employ, e.g., the Wilson score interval~\cite{Wilson:1927ug} to verify that our estimate fits well within the 99\% confidence interval of the corresponding estimate.}
\end{figure*} 

\section{Convergence analysis}
\label{app.Convergence}

In \cref{fig.Conv.ns.sample}, we provide details showing how our estimates of the relative volume converge as a function of $\Nt$, i.e., the number of sampled correlations.

\section{Distance from $\vecP_w$ to extreme points of $\NS$}
\label{app:DistanceOfExtremalPoints}

In the probability space $\sP$ parametrized by all the full conditional distributions $\{P(a,b|x,y)\}_{a,b,x,y}$, the Euclidean distance between two correlations $\vecP_1$ and $\vecP_2$ is given by:
\begin{equation}\label{eq:DistanceEuclidean}
  \De(\vecP_1|\vecP_2) = \sqrt{\sum_{a,b,x,y}\left[P_1(a,b|x,y)-P_2(a,b|x,y)\right]^2}. 
\end{equation}
In any Bell scenario, any Bell-local extreme point of $\NS$, which is also an extreme point of $\L$,  can be  obtained from 
\begin{equation}\label{Eq:Local-ExtPt} 
	P_{\L}(a,b|x,y) = \delta_{a,1}\delta_{b,1}
\end{equation} 
via a relabeling of measurement settings, outcomes, and/or parties. Since the uniform distribution $P_w(a,b|x,y) = \frac{1}{\no^2}$ is always invariant under such a relabeling, we see that their Euclidean distance is invariant under relabeling, and is easily shown to be $\De(\vecP_{\L}|\vecP_w)=\ns\sqrt{1-\frac{1}{\no^2}}$.

Similarly, in the $(2,\no)$ Bell scenarios, all nonlocal extreme points of $\NS$ can be obtained from one of the followings~\cite{Barrett_05}: 
\begin{equation}\label{Eq:NS-ExtPt} 
	 P_{\NS}^k(a,b|x,y) = \frac{1}{k}\delta_{(b-a)\, \text{mod}\, k,\, xy},\quad k \in {2,\dots, \no}
\end{equation}
via a relabeling. It then follows from~\cref{eq:DistanceEuclidean} that $\De(\vecP^k_{\NS}|\vecP_w)=2\sqrt{\frac{1}{k}-\frac{1}{\no^2}}$. Comparing with $\De(\vecP_{\L}|\vecP_w)$, we thus see that all nonlocal extreme points in the $(2,\no)$ Bell scenario are actually nearer to $\P_w$ than the local ones.

\bibliographystyle{plainnat}

\end{document}